\documentclass[useAMS,usenatbib]{mn2e}

\usepackage{times}
\usepackage{epsfig}
\usepackage{graphicx}
\usepackage{amsmath}	
\usepackage{natbib}
\usepackage{latexsym}
\newcommand{\msun}{M$_{\rm \odot}$}
\newcommand{{\kms}}{km\,s$^{-1}$}
\def\msunyr {M$_{\odot}~ \rm yr^{-1}$}

\graphicspath{{./IMAGES}}

\begin{document}
\renewcommand{\thefootnote}{\fnsymbol{footnote}}
\title[The many sides of RCW 86]{ The many sides of RCW 86: 
a type Ia supernova remnant evolving in its progenitor's wind bubble
}
\author[S. Broersen et al.]{Sjors~Broersen,$^1$ Alexandros~Chiotellis,$^1$ Jacco~Vink$^{1,2}$ and Aya~Bamba$^3$ \\ 
$^1$ Astronomical Institute `Anton Pannekoek', University of Amsterdam, Postbus 94249, 1090 GE Amsterdam, The Netherlands \\
$^2$ GRAPPA, University of Amsterdam, Postbus 94249, 1090 GE Amsterdam, The Netherlands \\
$^3$ Department of Physics and Mathematics Aoyama Gakuin University, Fuchinobe 5-10-1, Chuo-ku Sagamihara, Kanagawa, Japan, 252-5258}

\maketitle

\begin{abstract}

We present the results of a detailed investigation of the Galactic supernova remnant RCW 86 using the {\it XMM-Newton} X-ray telescope. RCW 86 is the probable remnant of SN 185 A.D, a supernova that likely exploded inside a wind-blown cavity. We use the {\it XMM-Newton} Reflection Grating Spectrometer (RGS) to derive precise temperatures and ionization ages of the plasma, which are an indication of the interaction history of the remnant with the presumed cavity. We find that the spectra are well fitted by two non-equilibrium ionization models, which enables us to constrain the properties of the ejecta and interstellar matter plasma. Furthermore, we performed a principal component analysis on EPIC MOS and pn data to find regions with particular spectral properties. We present evidence that the  shocked ejecta, emitting Fe-K and Si line emission, are confined to a shell of approximately 2 pc width with an oblate spheroidal morphology. Using detailed hydrodynamical simulations, we show that general  dynamical and emission properties  at different portions of the remnant can be well-reproduced by a type Ia supernova that exploded in a non-spherically symmetric wind-blown cavity. We also show that this cavity can be created using general wind properties for a single degenerate system. Our data and simulations provide further evidence that RCW 86 is indeed the remnant of SN 185, and is the likely result of a type Ia explosion of single degenerate origin. 
\end{abstract}

%\titlerunning{RCW 86.}
%\authorrunning{S. Broersen et al.}

\begin{figure}
\includegraphics[width=84mm]{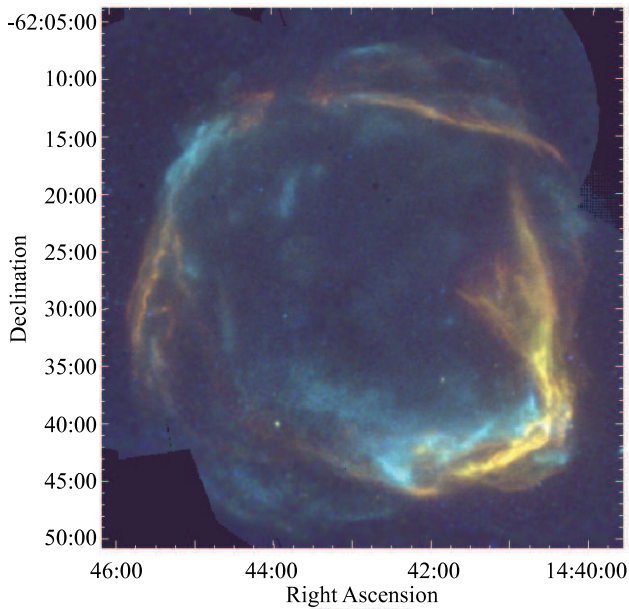}
\caption{A three colour image of RCW 86. The red, green and blue colours denote 0.5-1.0 keV, 1.0-1.95 keV and 2.0-5.0 keV emission, respectively.}
\label{fig:rgb}
\end{figure}

\section{Introduction}

Supernovae (SNe) chemically enrich and energise the interstellar medium (ISM). 
Part of their explosion energy is used in the supernova remnant phase to accelerate particles and in fact, SNRs are thought to be the main contributor to Galactic cosmic rays with energies up to the knee in the cosmic ray spectrum ($\approx10^{15}$ eV).  
Type Ia SNe are of interest as cosmological standard candles, and have been essential for the discovery
that the Universe is accelerating \citep{riessetal1998,perlmutteretal1999}. However, the progenitor systems of Type Ia supernovae are still a matter of debate \citep[e.g.][]{maozmanucci2012}.

Both the topic of particle acceleration in supernova remnants (SNRs) and the nature of type Ia supernovae make the SNR RCW 86 (also known as G315.4-2.3 or MSH 14-6{\it 3}) a very interesting object.
Although the name RCW 86 originally referred to the optically bright southwestern region, it is now also associated with the total remnant and we will therefore use it throughout this paper. It has been suggested that it is the remnant of an event recored by Chinese astronomers in the year 185 A.D. \citep{clarkandstephenson1975}, although this is still a matter of debate  \citep[see e.g.][]{dickeletal2001, smith1997, vinketal2006}. 
Located at a distance of $2.5 \pm 0.5$~kpc  \citep[][]{helderetal2013}\footnote{
For a long time it was unclear whether RCW 86 is located at a distance of $\sim 2.5$~kpc
\citep{westerlund1969,rosadoetal1996}, or much more nearby at $\sim 1$~kpc \citep{longblair1990,bocchinoetal2000}. The recent proper motion measurements of
\citet{helderetal2013}, combined with plasma temperature measurements based
on the broadline H$\alpha$ emission, now clearly
 indicates that RCW 86 is at a distance of 2.5~kpc, or even further
if the plasma temperature is quenched due to cosmic-ray acceleration.
}, RCW 86 is a shell-type SNR with an angular diameter of approximately 40 armin, making it unusually large (R $\approx 15d_{2.5}$ pc) for its age,
if it is indeed the remnant of SN 185. For the remnant to have reached this size in 1830 years, it must have been expanding with a mean velocity of around 7800 km~s$^{-1}$. This high velocity, but also several other characteristics, have led to the suggestion that the SNR has been expanding in a low density, wind-blown cavity \citep*[]{vinketal1997}. The measured expansion velocities of 500-900 km s$^{-1}$ in the SW and NW \citep{longblair1990,ghavamianetal2001}, and the $\approx$1200~km~s$^{-1}$ measured in the NE and SE portions of the remnant \citep{helderetal2013} suggest that different parts of the remnant are in different stages of interaction with the dense surroundings of the wind cavity.

X-ray images of RCW 86 reveal a non-spherically symmetric shell with different morphologies in the
soft (0.5-2 keV) and hard (2-5 keV) X-ray bands \citep*{vinketal1997}, as illustrated in  Fig. \ref{fig:rgb}. 
\citet{rhoetal2002} found, using Chandra data, that the hard X-ray emission in the south western part of the remnant is close to an Fe-K line emitting region. They suggest that the hard X-ray continuum is synchrotron radiation coming from electrons accelerated at the reverse shock of the remnant. 
Besides the non-thermal emission in the SW, there is also synchrotron emission present in the NE (\citet*{bambaetal2000}, \citet{vinketal2006}) and, somewhat fainter, in the NW \citep{yamaguchietal2011, williamsetal2011, castroetal2013}.
X-ray synchrotron radiation requires the presence of 10-100 TeV electrons, the presence of which has been corroborated
since then by the detection of TeV gamma-ray emission from this remnant \citep{aharonianetal2009,rcw86cosmicrayconstraints}. In addition, the amplification as observed by \citet{vinketal2006} and \citet{castroetal2013} also suggests efficient particle acceleration at the shock of RCW 86. 
The measured shock velocities in the optical of  600--1500 km s$^{-1}$, however, are too low to explain the occurrence of X-ray synchrotron emission 
\citep[e.g.][]{Zirakashviliaharonian2007}. In this regard, RCW 86 differs from the other young Galactic SNRs Cas A, Kepler, Tycho and SN 1006, for which X-ray synchrotron emission is accompanied by measured shock velocities in the range of 3000--5400 km s$^{-1}$.
\citet{helderetal2013} argued for the NE region that either the shock velocity was much higher in the recent past (and the shock slowed down on a timescale much shorter than the synchrotron cooling time
of the electrons), 
or, as also argued by \citet{castroetal2013}, the shock velocity measured in Balmer line emitting shocks are lower than those of synchrotron emitting shocks. This is supported by the higher shock
velocity measurement in X-rays for the northeastern part of the SNR ($V_s = 6000$ km s$^{-1}$ \citet{helderetal2009}).
The latter possibility could arise if the cavity wall exists of clumpy material, and the H-alpha shocks arise from parts of the forward shock which are moving more slowly, in higher density regions.

\citet{uenoetal2007} used the Suzaku telescope to map the Fe-K emission in the southwestern part of the remnant. They found that the distribution of
the Fe-K line emission anti-correlates with the hard X-ray continuum (3.0-6.0 eV), but  that in fact the Fe-K emission correlates well with the radio synchrotron emission. Since radio synchrotron emission originates from regions somewhat downstream of the forward shock, they conclude that the Fe-K emission must come from shocked ejecta. In addition, they measured an intrinsic line broadening in this ejecta component of $\approx50$ eV. \citet*{yamaguchietal2011} used additional Suzaku observations to take a more detailed look at the Fe-K emission in the whole of RCW 86. They confirmed that the line centroid suggests a low ionization state of Fe and suggest a type Ia progenitor based on the amount of Fe present in the remnant. 

The question of what the type of supernova is that led to the formation of RCW 86 is still open. The remnant is located in close vicinity to several B-type stars, which suggests RCW 86 is the result of a core-collapse supernova \citep{westerlund1969}. Recently, however, \citet{williamsetal2011} argued strongly that it is the remnant of a type Ia explosion, pointing to
1) the all-around presence of Balmer filaments \citep{smith1997}, 2) the high Fe mass found in the interior of the remnant, 3) the lack of a pulsar wind nebula or neutron star in the SNR, and 4) the lack of high abundance O emitting regions.
They also show, using hydrodynamical simulations, that if RCW 86 is indeed the remnant of SN 185 A.D., the currently observed ambient medium densities, expansion velocities and size can only be explained if the remnant expanded in a low-density cavity. 
If RCW 86 is the remnant of a type Ia explosion, a cavity can be created by a high velocity accretion-wind \citep*{hachisuetal1996, badenesetal2007}, which requires a white dwarf that accretes material in a rate higher than the critical rate for stable hydrogen burning. A confirmation of the SN explosion type for RCW 86 would therefore not only be a confirmation that type Ia supernovae can arise through the single degenerate channel, but also that these progenitor systems can actively modify their environment. 

In this work we aim to investigate the issues outlined above using the {\it XMM-Newton} X-ray telescope. We use the high spectral resolution of the RGS instrument to investigate the interaction history of the remnant with the cavity wall, and the imaging and spectral capabilities of the EPIC CCDs to have a more detailed look at the presence of Fe-K and other ejecta emission close to the forward shock. In addition, we use the principal component analysis (PCA) technique to highlight regions of interest, which we then further investigate using the EPIC instrument. Finally, we use hydro-simulations to show that the size, the dynamics and the emission properties of RCW 86 can be well-reproduced by a single degenerate wind-blown cavity scenario. 

\begin{table}
\caption{List of {\it XMM-Newton} observations of RCW 86.}
\label{tab:po}
\begin{center}
\begin{tabular}{lllrr}
OBSID & RA & DEC & time [s] & orbit \\
\hline

0110011301 & 220.56 & -62.37 & 19566 & 126 \\ 
0110011401 & 220.51 & -62.22 & 18677 & 126 \\ 
0110010701 & 220.73 & -62.63 & 23314 & 126 \\ 
0110010501 & 220.14 & -62.60 & 16097 & 309 \\ 
0110012501 & 220.24 & -62.72 & 12232 & 592 \\ 
0208000101 & 221.26 & -62.30 & 59992 & 757 \\ 
0504810101 & 221.57 & -62.30 & 116782 & 1398 \\ 
0504810601 & 221.57 & -62.30 & 36347 & 1399 \\ 
0504810201 & 221.40 & -62.47 & 75216 & 1406 \\ 
0504810401 & 220.15 & -62.60 & 72762 & 1411 \\ 
0504810301 & 220.50 & -62.22 & 72381 & 1412 \\ 

\hline
\end{tabular}
\end{center}
\end{table}

\section{Data Analysis}

\subsection{XMM-Newton data}

For our analysis, we used all {\it XMM-Newton} pointings available (see Table \ref{tab:po}) for RCW 86. All of the listed EPIC data were used to create Fig. \ref{fig:rgb}, using the extended source analysis software ESAS \citep*{kuntzsnowden2008}.  

\begin{figure}
\includegraphics[width=64mm]{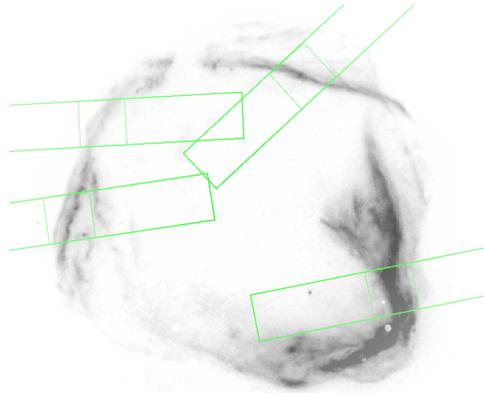}
\caption{An inverted greyscale image of the 0.5-1.0 keV band of RCW86. Overlaid in green are the regions from which the RGS and MOS spectra shown in section \ref{sec:rgs} are extracted. }
\label{fig:rgs_regions}
\end{figure}

For the Reflection Grating Spectrometer \citep{denherderetal2001}, we used the pointings shown in Fig. \ref{fig:rgs_regions}. The data were screened for flaring, after which we extracted the spectra using the normal pipeline software SAS V 12.0.1. Because RCW 86 is an extended source, the edges of the RGS CCDs cannot be used for background subtraction. We therefore used blank sky observations of similar orbit and large observation time as background. 
The often extended angular size of supernova remnants can be a problem when observing with the RGS. It causes emission lines to broaden, for photons of the same wavelength enter the instrument at slightly different angles, and are therefore reflected onto the CCDs at slightly different angles. Although RCW 86 has a large angular diameter, the line broadening is not such a problem as it is in e.g. SN 1006 \citep{broersenetal2013}, which has similar angular diameter. RCW 86 has a shell-like structure, with the thermal emission located mostly in narrow filaments of only a few arc minutes in width. Since the line broadening is approximately given by $\Delta \lambda = 0.124 \left(\frac{\Delta \phi}{1'}\right)$ \AA, where $\Delta \phi$ is the angular width of the emitting region, the broadening is limited to a few tenths of \AA. We correct for this effect, as in previous work \citep[][]{broersenetal2013}, by convolving the RGS response matrix with the emission profile along the dispersion axis. 

For the EPIC spectra we use the normal XMM-SAS pipeline software. The data were screened for periods of high flaring using the ESAS software. The spectral analysis was done with SPEX version 2.03.03 \citep*{spex}. The errors are 1$\sigma$ unless otherwise stated. 

\begin{table}
\caption{Energy bands of images created as input for the principal component analysis.}
\label{tab:pcbands}
\begin{center}
\begin{tabular}{r@{~--~}lll}
\hline
\multicolumn{2}{c}{Energy}  & \multicolumn{2}{c}{spectral}   \\
\multicolumn{2}{c}{$[\rmn{eV}]$} & \multicolumn{2}{c}{association} \\

\hline
500 & 600 & \mbox{O\, {\sc vii}} &  \\
601 & 700 & \mbox{O\, {\sc viii}} & \\ 
701 & 860 & \mbox{Fe {\sc - L}} &\\
861 & 980 & \multicolumn{2}{l}{\mbox{Ne\, {\sc ix}} He-$\alpha$ / \mbox{Fe {\sc - L}}} \\
981 & 1250 &\multicolumn{2}{l}{\mbox{Ne\, {\sc ix}} He-$\beta$ / \mbox{Fe {\sc - L}}}   \\
1251 & 1500 & \mbox{Mg\, {\sc xi}}& \\
1701 & 1999 &  \multicolumn{2}{l}{\mbox{Si {\sc vi-xiii}}} \\
2000 & 5000 & \multicolumn{2}{l}{Continuum}  \\
6300 & 6500 & \mbox{Fe\, {\sc -K}} &\\
\hline

\end{tabular}

\end{center}
\end{table}

\subsection{Principal component analysis}
\label{sec:pca_data_analysis}

As mentioned above, we used principal component analysis \citep[e.g.][for an extended description of the subject]{pcabook} to find regions of interest for spectral extraction. Principal component analysis (PCA) is a statistical analysis technique which was successfully applied by \citet{warrenetal2005} for the analysis of Tycho's SNR. 

The general goal of PCA is to reduce the dimensionality of a dataset consisting of a large number of variables,
while preserving as much as possible the variation (i.e. information) present in the dataset.
This is achieved by transforming a number of $n$ original variables, which likely contain internal correlations,
to a new set of uncorrelated variables (principal components, PCs),
which are ordered based on the amount of variation they represent of the original dataset. 
In general the first $m<n$ PCs will account for most of the variation in the data set.
As input variables we use $n$ images in the energy bands listed in Table \ref{tab:pcbands}. 
Our output images (or PCs) are linear combinations of the original input images.
If our original images are labeled $\overline{X}_i$, then our output PC images
are $\overline{Y}_j=a_{ij}\overline{X}_i$, with the coefficients $a_{ij}$ forming an orthogonal basis, with lower
values of $j$ accounting for more of the variance in the original data set.
The technique is blind in the sense that the output PCs do not necessarily have a straightforward physical interpretation, but nonetheless PCA is useful tool for finding correlations in the data. 
We show part of the results of this analysis in section \ref{sec:pca}, the full results are available online.

\begin{figure}
\includegraphics[width=84mm]{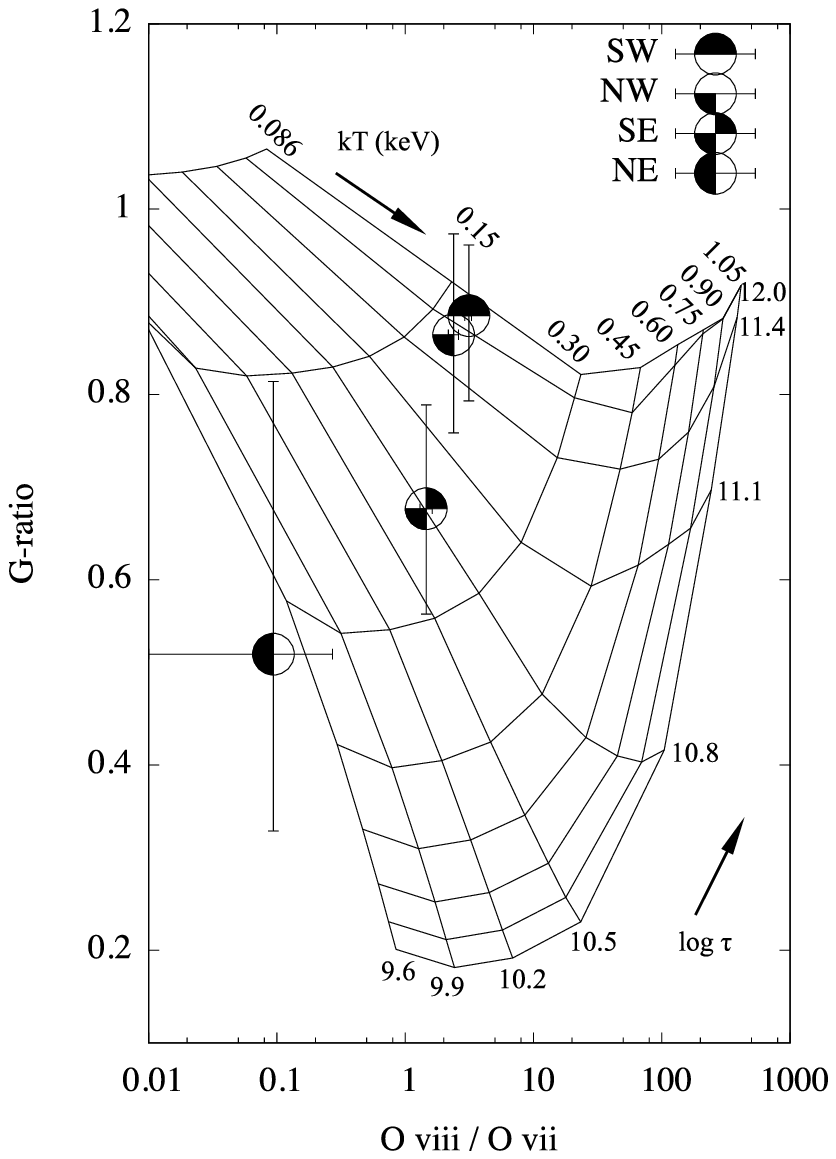}
\caption{Grid of temperature and n$_{\rmn e}$t ($\tau$) values created with SPEX, with the observed line ratios plotted of the different regions of the remnant. This figure is inspired by the final figure in \citet{vedderetal1986}. Note that, although this is not shown in the figure, the G-ratio increases dramatically for $\tau \la 2\times 10^{9}$ cm$^{-3}$ s \citep{vink2012}, while the \mbox{O\, {\sc viii}} / \mbox{O\, {\sc vii}} ratio keeps decreasing, so that the NE region can also be fit by NEI models with high $kT$ and low $\tau$.}
\label{fig:o_grid}
\end{figure}

\begin{figure*}%[htbp]       & trim l b r t
\begin{center}$
\begin{array}{cc}
\includegraphics[trim=0 -10 -50 0,clip=true,width=0.5\textwidth,angle=0]{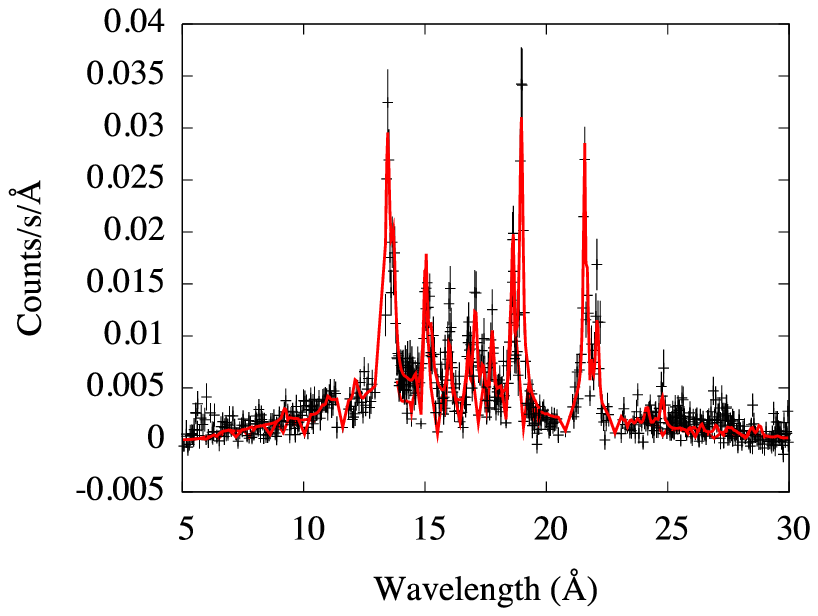} &
\includegraphics[trim=0 -10 -50 0,clip=true,width=0.5\textwidth,angle=0]{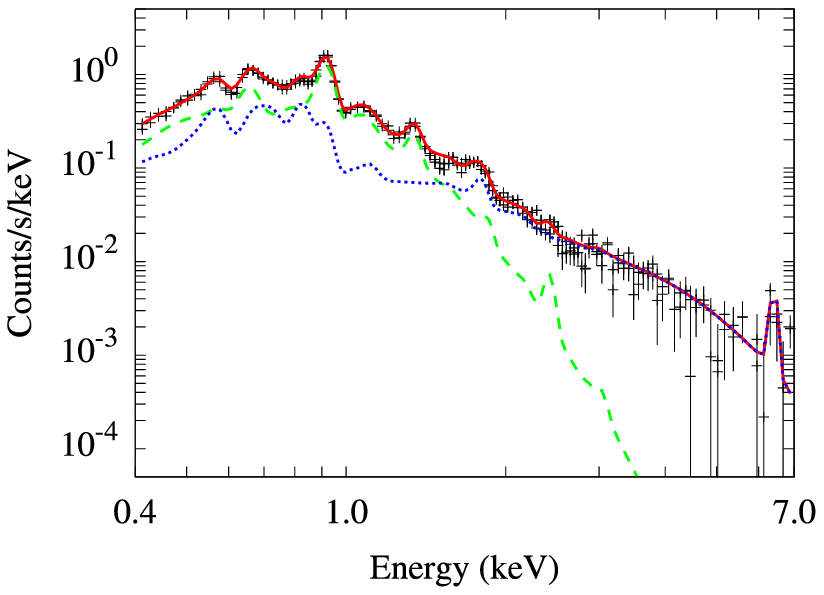} \\
\includegraphics[trim=0 -10 -50 0,clip=true,width=0.5\textwidth,angle=0]{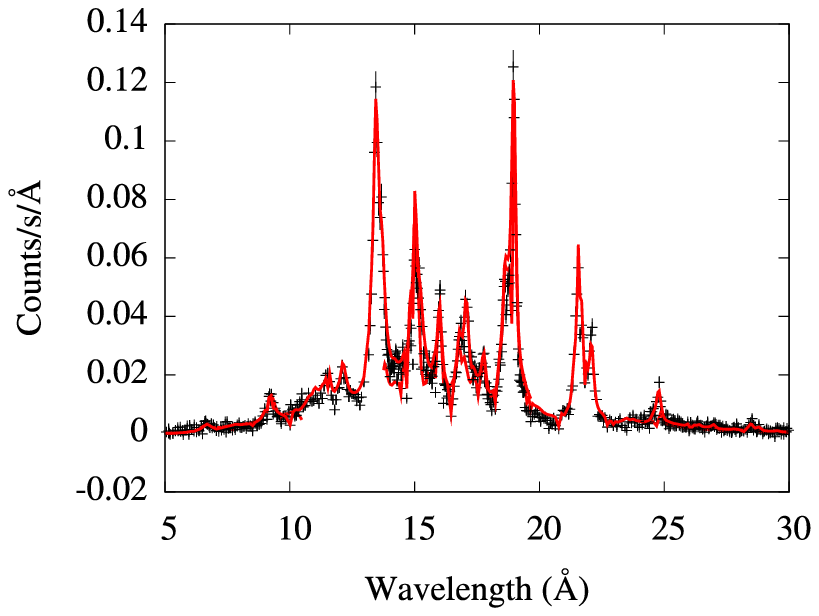}  &
\includegraphics[trim=0 -10 -50 0,clip=true,width=0.5\textwidth,angle=0]{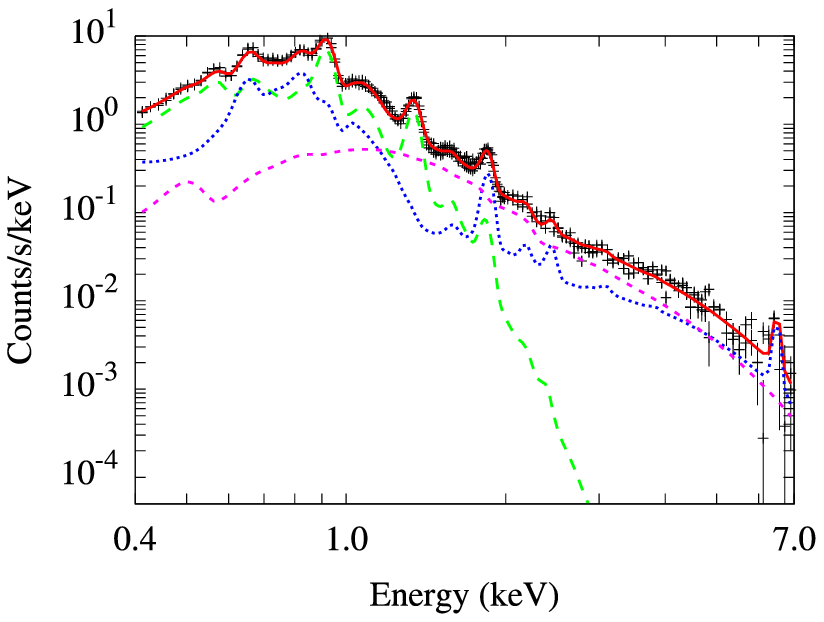} \\
\includegraphics[trim=0 -10 -50 0,clip=true,width=0.5\textwidth,angle=0]{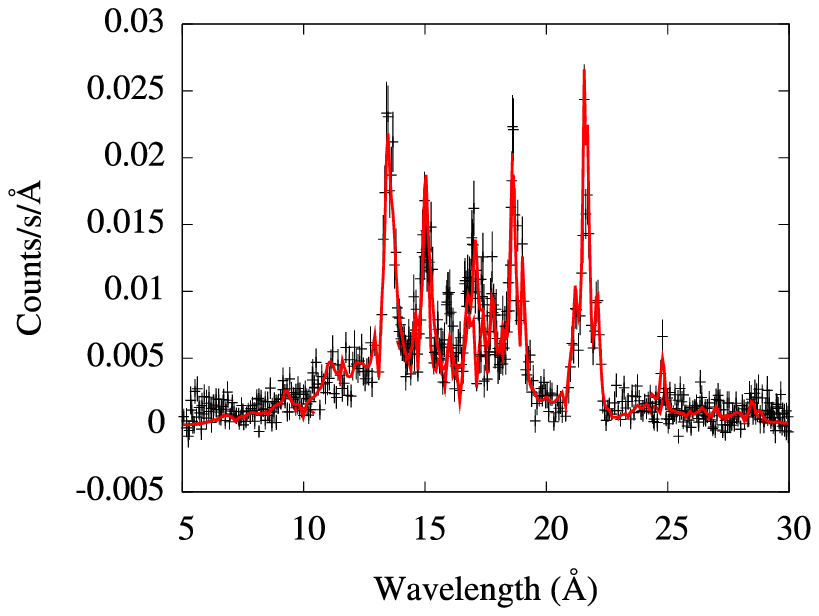} &
\includegraphics[trim=0 -10 -50 0,clip=true,width=0.5\textwidth,angle=0]{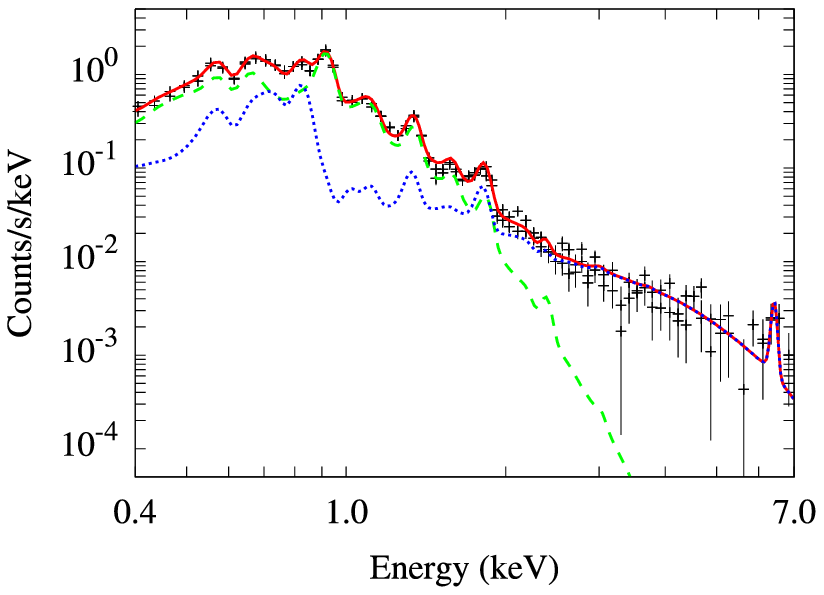}\\
\includegraphics[trim=0 -10 -50 0,clip=false, width=0.5\textwidth,angle=0]{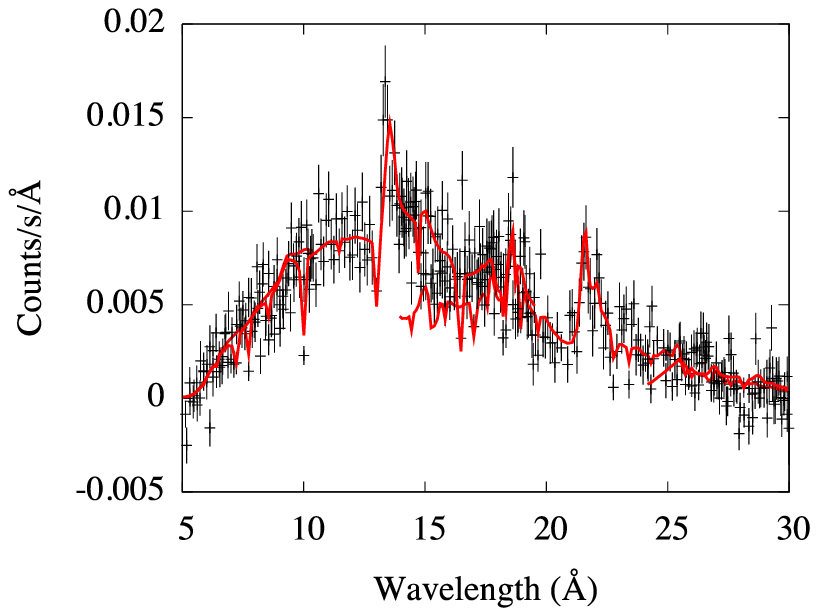}  &
\includegraphics[trim=0 -10 -50 0,clip=false,width=0.5\textwidth,angle=0]{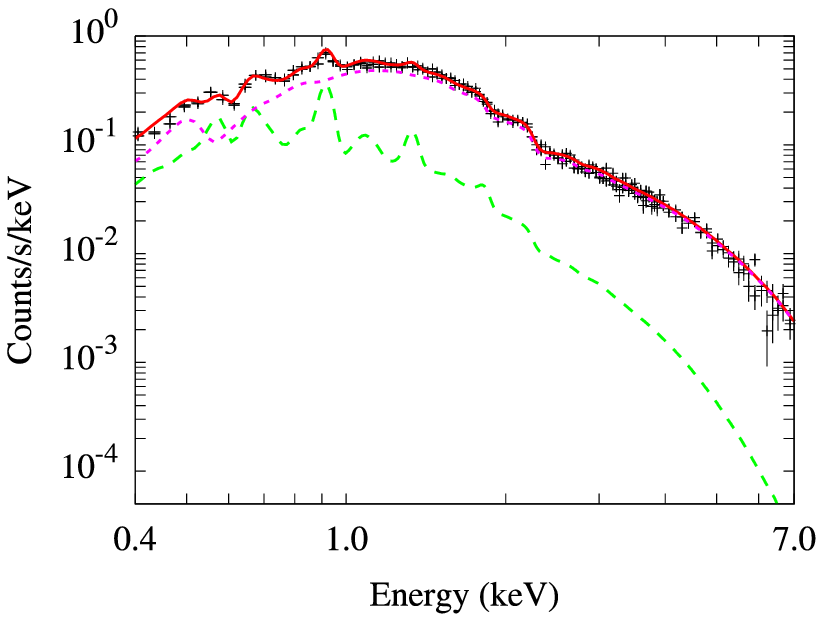} 
\end{array}$
\end{center}
\caption{The RGS (left) and MOS (right) spectra of the, from top to bottom, NW, SW, SE and NE region of RCW 86. In the MOS spectra, the solid red line represents the best fit model with the parameters shown in Table \ref{tab:spectral_parameters} , which consists of a low $kT$ (green, long dash) and a high $kT$ (blue, dotted) NEI component, and in the SW and NW region a powerlaw (magenta, dashed). 
  \label{fig:spectra}}
\end{figure*}

\begin{figure}
\includegraphics[width=64mm,angle=-90]{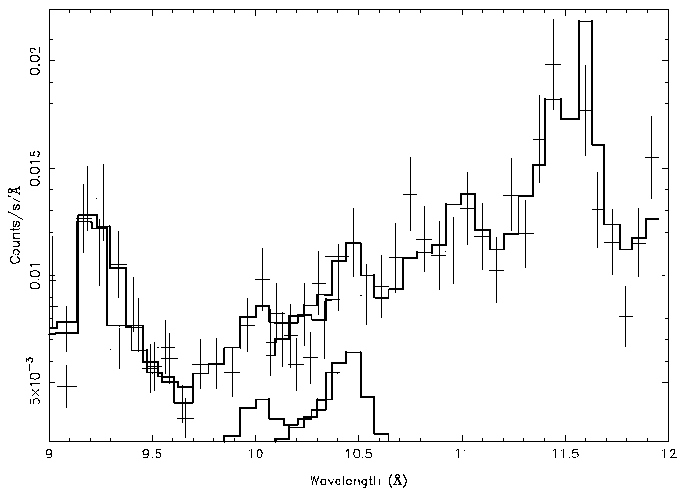}
\caption{Close up of the 9-12 \AA~region. The model (solid black line) consists of a single NEI model combined with two Gaussians, which are also shown. Both RGS 1 and RGS 2 data and model are shown. Note that the RGS 1 region has a chip defect, due to which it cannot detect emission in the 10--14 \AA~range. }
\label{fig:9_12_ang}
\end{figure}

\begin{table*}
\begin{minipage}{126mm}
\caption{ Results of the joint MOS (0.4-7.0 keV band) and RGS (5-30 \AA\ band) fitting of different regions of the remnant. }
\label{tab:spectral_parameters}

\begin{tabular}{llcccc}
\rule{0pt}{5mm}
%&  & \multicolumn{4}{c}{Region} \\
%\cline{3-6}
%\multicolumn{5}{l}{$\ion{O}{VII} $} \\
 Parameter & Unit & SW &  NW & SE & NE \\ 
\hline

$N_{\rmn H}$ &$10^{21}$ cm$^{-2}$	&$4.80_{-0.07}^{+0.12}$	& $3.64_{-0.11}^{+0.11}$ & $3.81_{-0.09}^{+0.08}$ & $3.67_{-0.05}^{+0.06}$ \\ 
\rule{0pt}{5mm}
\hspace{-0.8mm}ISM & & & & & \\
$n_{\rmn e}n_{\rmn H}V$ &$10^{58}$ cm$^{-3}$  & $1.17_{-0.16}^{+0.22}$ &$0.16_{-0.02}^{+0.03}$ & $0.16_{-0.01}^{+0.02}$ & $0.50_{-0.05}^{+0.08}$ \\ 
$kT$ & keV					& $0.19_{-0.01}^{+0.01}$	 	&$0.34_{-0.02}^{+0.02}$ 		& $0.32_{-0.01}^{+0.01}$ & $1.16_{-0.22}^{+019}$ \\ 
$\tau$ &$10^{9}$ cm$^{-3}$ s 	&$90.5_{-9.9}^{+14.7}$		 &$62.5_{-10.7}^{+15.8}$ 		& $29.1_{-1.1}^{+1.1}$ 	& $4.28_{-0.35}^{+0.54}$ \\ 

C$^\dag$\footnotetext{$^\dag$The abundances are in terms of $N_{\rm x}$ / $N_{\rmn{x, solar}}$, where $N_{\rm x}$ is the number of particles of the element in question. }

			&			&$0.65_{-0.12}^{+0.21}$		&$0.42_{-0.21}^{+0.32}$		& $0.57_{-0.13}^{+0.12}$ 	& $0.2$\\ 
N &						&$0.40_{-0.04}^{+0.07}$		&$0.05_{-0.02}^{+0.02}$ 		& $0.19_{-0.03}^{+0.03}$ 	& $0.2$ \\ 
O &						&$0.42_{-0.03}^{+0.05}$		&$0.12_{-0.01}^{+0.02}$ 		& $0.19_{-0.04}^{+0.03}$ 	& $0.2$ \\ 
Ne &	 					&$0.84_{-0.06}^{+0.11}$		&$0.22_{-0.03}^{+0.03}$ 		& $0.26_{-0.02}^{+0.01}$ 	& $0.2$ \\ 
Mg  &					&$1.16_{-0.08}^{+0.17}$		&$0.13_{-0.01}^{+0.02}$ 		& $0.18_{-0.03}^{+0.03}$ 	& $0.2$ \\ 
Si&						&$0.40_{-0.07}^{+0.09}$		&$0.08_{-0.03}^{+0.03}$		& $0.34_{-0.10}^{+0.11}$ 	&$0.2$ \\
Fe &						&$0.35_{-0.05}^{+0.12}$		&$0.02_{-0.01}^{+0.01}$ 		& $0.06_{-0.01}^{+0.01}$ 	& $0.2$ \\ 
\rule{0pt}{5mm}

\hspace{-0.8mm}Ejecta  & & & & & \\
$n_{\rmn e}n_{\rmn H}V$ &$10^{58}$ cm$^{-3}$	&$1.98_{-0.19}^{+0.30}(\times10^{-3})$	&$6.76_{-0.53}^{+0.48}(\times10^{-3}) $ & $2.0_{-0.4}^{+0.5} (\times10^{-3})$ & $-$ \\ 
$kT$ &keV 					&$3.23_{-0.10}^{+0.12}$		&$2.43_{-0.19}^{+0.23}$ 		& $3.46_{-0.33}^{+0.44}$ 	& $-$ \\ 
$\tau$ &$10^{9}$ cm$^{-3}$ s 	&$15.5_{-0.03}^{+0.03}$		&$2.00_{-0.06}^{+0.07}$ 		& $2.69_{-0.11}^{+0.21}$ 	& $-$ \\ 
C &						& $1$					&$1$ 					& $1$ 				& $-$ \\ 
N &						&$1$					&$1$ 					& $1$ 				& $-$ \\ 
O &						&$17.9_{-2.0}^{+2.1}$		&$0.71_{-0.12}^{+0.11}$ 		& $1.61_{-0.29}^{+0.33}$ 	& $-$ \\ 
Ne &						&$3.54_{-0.52}^{+0.44}$		&$0.26_{-0.11}^{+0.12}$ 		& $0.00_{-0.00}^{+0.14}$ 	& $-$ \\ 
Mg &						&$1$					&$1$ 					& $0.39_{-0.21}^{+0.19}$ 	& $-$ \\ 
Si &						&$1.78_{-0.20}^{+0.19}$		&$0.85_{-0.16}^{+0.17}$ 		& $1.37_{-0.35}^{+0.32}$ 	& $-$ \\ 
Fe &						&$3.38_{-0.44}^{+0.30}$		& $23.8_{-6.1}^{+8.2}$ 		& $7.53_{-3.01}^{+2.90}$ 	& $-$ \\ 
\rule{0pt}{5mm}
\hspace{-0.8mm}Power law  & & & & & \\
norm &$10^{44}$ ph s$^{-1}$ keV$^{-1}$	&$1.45_{-0.04}^{+0.08}(\times10^{-2})$		 &$-$ & $-$ 				& $0.69_{-0.01}^{+0.01}$ \\ 

$\Gamma$$^\ddag$\footnotetext{$^\ddag$ The photon number index, i.e.: $N(E) \propto E^{-\Gamma}$.}

		&				&$3.82_{-0.05}^{+0.09}$		&$-$ 				& $-$ 				& $2.67_{-0.02}^{+0.02}$ \\ 

\hline
C-stat / d.o.f.&				&3336 / 1021  				&1961 / 986 			&2031 / 902 		& 1172 /  709

\end{tabular}
%\footnotetext{$^1$ }
%\footnotetext{}  
\end{minipage}
\end{table*}

%\begin{figure}
%\includegraphics[width=64mm,angle=-90]{IMAGES/0401_mos_rgs_final_spec.eps}
%\caption{The RGS + MOS spectrum of the SW part of the remnant. }
%\label{fig:rgs_SW}
%\end{figure}

\section{Results}
\subsection{RGS Data}
\label{sec:rgs}

RCW 86 has been observed using {\it XMM-Newton} on numerous occasions (see Table \ref{tab:po}).
We focus here on four RGS observations, which were chosen such that each of the four quadrants of the remnant were covered. 

The high resolution of the RGS allows for accurate measurements of fluxes and energies of line emission, which in turn can be used as diagnostic tools for the state of the plasma from which it originates. When a plasma is suddenly shocked it is under-ionised, after which it will go to ionization equilibrium on a density dependent timescale $t \simeq 10^{12.5}/ n_{\rmn e}$ s, where $n_{\rm e}$ is the electron density. The quantity $\tau =  n_{\rm e}t$ is referred to as the ionization age \citep{vink2012}. This ionization age, together with the temperature $kT$, are the most important parameters in non-equilibrium ionization (NEI) models which have been successfully applied in describing SNR plasmas. 
An important plasma diagnostic is the G-ratio, which is a measure of both temperature and ionization age $\tau$ \citep[e.g.][]{vink2012}. 
This G-ratio is based on the He-like line triplet of $\mbox{O\, {\sc vii}}$, and is defined as (F+I) / R, where  
F ($\lambda$=22.098 \AA) $1s2s^3 S_1 \rightarrow 1s^{2}~ ^{1}S_0$ is the forbidden transition, I ($\lambda$ = 21.804, 21.801 \AA) is the sum of the two intercombination transitions $1s2 p^3 P_{1,2} \rightarrow 1s^{2}~ ^{1}S_0$, and R ($\lambda$ = 21.602 \AA) is the resonance transition $1s2p ^{1}P_1\rightarrow 1s^{2}~ ^{1}S_0 $. 
A different diagnostic is based on the ratio of $\mbox{O\, {\sc viii}}~ \rm He\alpha$ / $\mbox{O\, {\sc vii}}~ \rm He\beta$, 
which can be related to the ionization age. 
To measure the strengths of the triplet lines, we first 
fitted the spectrum with a single absorbed NEI model.
We then used this to model the overall spectrum, except for the
triplet lines themselves. For the triplet lines we
used gaussian line profiles, with the energies fixed
to the laboratory values. The line widths were allowed to vary,
but were coupled for the lines. A similar procedure was used for
all emission lines listed in Table~\ref{tab:lines}. Thos table shows the line strengths of the O lines and of $\mbox{Fe\, {\sc xvii}}$ lines, which can also be used as a diagnostic \citep{gillaspyetal2011}. The line ratios of O  can be combined to provide a unique measurement of both $kT$ and $\tau$ of a plasma, based solely on  lines which are often very prominent in SNR plasmas. As in \citet{vedderetal1986} we have created a grid of the two O based line ratios using the NEI model in SPEX, in which grid we have plotted the measured line ratios Fig.~\ref{fig:o_grid}.

Although the error bars are quite large, the line ratios still show a clear trend in both ionization age and temperature. The SW and NW regions seem to have the highest ionization ages and lowest temperatures, while the SE and NE regions have higher temperatures and lower ionization ages. Although it is not shown in the figure, the G-ratio increases dramatically for high temperatures at ionization ages below $log(\tau) = 9.6$, so that the NE region line ratios are also consistent with a high $kT$, low $\tau$ model. 
The O lines in the RGS spectrum may be the result of a mix of different plasmas, but this still gives an overall correct trend of the $kT$ / $\tau$ of the O plasma, as confirmed by our spectral modelling (see Table \ref{tab:spectral_parameters}). 

We fitted the RGS spectrum simultaneously with the EPIC MOS 1 and 2 spectra, extracted from a square region of $5'\times5'$  lying exactly on the RGS pointing. 
The reason for this is that with the RGS data alone it is not possible to accurately determine the $kT$ of the high temperature model,  as the energy range of the instrument (0.3-2 keV, 5-40 \AA), makes it unsuitable to reliably  fit  continuum emission.
In order to make simultaneous fits, the MOS 
spectra were normalised to the RGS spectra, which is needed to take into account the effects of differences in the size of regions contributing to the RGS spectra. Fitting data from the two instruments together may artificially inflate the obtained C-stat / d.o.f., due to cross-calibration errors in the effective areas of the instruments. We used C-stat as a fit statistic \citep{cash},  for it is more reliable at lower count rates, and approaches $\chi^2$ for higher count rates. 
The RGS and MOS spectra are shown in Fig.~\ref{fig:spectra}.
We first attempted to fit the spectrum with the hydrogen column density $N_{\rmn H}$ as a free parameter combined with a single NEI model with fixed abundances. 
In all quadrants, the C-stat improved significantly both by freeing abundance parameters and by adding another NEI component to account for the presence of a different $kT$ plasma, or by adding a power law in the case non-thermal emission is expected.
Although the plasma properties vary between the different regions of the remnant (Table \ref{tab:spectral_parameters}) the models are largely consistent with each other, with the exception of the northeastern (NE) region.
In the southwest (SW), northwest (NW) and southeastern (SE) regions, the best fit model consists of two NEI components, of which one has sub solar abundances, low $kT$ and high $\tau$, while the other NEI component has high $kT$, lower  $\tau$, and elevated, super solar  abundances, most notably of Fe. In the SW region an additional power law is required to fit the data. This is consistent with figure \ref{fig:rgb}, where the brightest non-thermal emission is visible in the SW part of the remnant.
The elemental abundances of the low $kT$ model are significantly sub solar, with a mean value of $\sim0.2$. \citet{borkowskietal2001} already found that the continuum was strong and fixed the abundances at 2/3 of solar value. \citet{yamaguchietal2011} find abundance values which are also sub solar, with the exception of Ne in some regions.
Given that the NEI component with the higher temperature and low $\tau$ has
elevated abundances, it is likely that this component is associated
with shocked ejecta. The low $kT$ component is probably shocked ambient
medium. Our hydrodynamical modelling is consistent with this interpretation
(Section~\ref{sec:simulations}).

The SW region is the brightest region in the remnant both in thermal and non-thermal emission, as is clear from Fig. 1. This brightness results in a rich RGS spectrum (see Fig. \ref{fig:spectra}), which contains a plethora of emission lines from many different elements, including C, N, O, Ne, Mg, Si and Fe. The best-fit model of this region includes two NEI models with ionization age and $kT$ as mentioned above, and a power law. The figure shows that the model spectrum gives a qualitatively good fit. 
However, the  C-stat / degrees of freedom (d.o.f.) is in some cases above 2, which is formally unacceptable. 
The good statistics and high spectral resolution of the instrument make that the C-stat is more sensitive to systematic errors of instrumental nature and small inaccuracies and incompleteness of the atomic database.
Although the
remaining instrumental cross-calibration problems affect the C-stat, they
do not affect the overall best fit parameters. 
This is illustrated by the fact that when we fitted the same models separately  to the 
RGS and MOS spectra, it gives much better values
for C-stat / d.o.f. statistics for the individual spectra, 
with only minor differences in the best fit parameters compared to the joint
fits.
Overall, the main discrepancies between the model and the spectra appear to be the \mbox{Fe\, {\sc xvii}} lines at 15 \AA~and 17 \AA\ (1.21-1.37 keV), a region between 10 \AA~ and 12 \AA\ (1.24 - 1.03 keV) and the region between 19--21 \AA\ (1.53-1.69 keV). The difference between fit and model of the \mbox{Fe\, {\sc xvii}} lines seems mainly to originate from a normalisation difference between the both RGS and the MOS instruments. The spectrum between 10-12 \AA~ has been a problematic region to fit for a long time for different SNRs \citep[e.g.][]{broersenetal2011}, a problem which has mainly been attributed to missing Fe lines in the current plasma codes \citep[e.g.][]{bernittetal2012}. The good statistics of this spectrum give us the opportunity to identify which lines exactly are missing  in this case. The data require two additional Gaussians with wavelengths 10.04 \AA~and 10.4 \AA\ (0.81 and 0.84 keV, see Fig. \ref{fig:9_12_ang}). 
According to the NIST database there are several possible ions which could be responsible for these lines, including  \mbox{Ni\, {\sc xxi}} and  \mbox{Fe\, {\sc xxi}}, or higher charge states of these elements. There are, however, \mbox{Fe\, {\sc xviii}} lines present at both wavelengths. Since, for the NEI parameters found, most of the Fe has not been ionised to \mbox{Fe\, {\sc xx}}, \mbox{Fe\, {\sc xviii}} seems the most likely candidate to account for the missing lines in this region. 

The overall SW spectrum is well fit by our two component model. However, our model predicts a higher centroid energy
for the Fe-K emission in this region than observed. The Fe-K emission in this region, as elsewhere in the SNR, has a centroid consistent with 6.4~keV, consistent with $\tau < 2\times10^{9}$ cm$^{-3}$ s, but
in our model it is fit with a component with an ionization age of $1.5\times 10^{10}$cm$^{-3}$s. Likely the two component model for this
region is an oversimplification, but unfortunately a three-component model is ill-constrained.

For the NW and SE region, the fit parameters of the ISM, low $kT$ component are very similar, with the exception of the ionization age. The high $kT$ NEI component accounts for most of the Fe and Si and some of the \mbox{O\, {\sc vii}} emission. The centroid energy of the Si-K line is a direct measurement of the ionization state of the plasma, where a higher centroid indicates further ionised Si. The low centroid of Si-K at 1.80 keV in this region cannot be accounted for by the ambient medium NEI component, because its ionization age is too high. The centroid energy and ionization age make it likely that the the Si-K and the Fe-K originate from the same ejecta plasma, which is expected to have been shocked more recently by the reverse shock. The principal component analysis detailed in section \ref{sec:pca} indicates that the Si and the Fe ejecta are co-located, and
also suggests that they originate from the same plasma. Fitting Si-K and Fe-K with  a single NEI model seems therefore justified.
The $kT$ and $\tau$ of the ambient medium component are a bit different than expected from Fig. \ref{fig:o_grid}, which is caused by a significant contribution to the \mbox{O\, {\sc vii}} emission by the hottest component (Fig.~\ref{fig:spectra}). 

The NE region of the remnant is mostly dominated by non-thermal emission and therefore the RGS spectrum shows weak lines only of O and Ne. The best fit model contains both a power law and an NEI component with abundances fixed at 0.2 times solar, more or less the mean value obtained from fits of other regions. Contrary to the other three regions, this region does not show emission associated with ejecta components. As is already apparent from Fig. \ref{fig:o_grid}, there is a degeneracy in the model for the NE region, in that the data can be fit both by a high $kT$, low $\tau$ model, or a low $kT$, $log(\tau) \approx 10$ model. This became apparent while fitting the data as well, in that both cases of the model gave very similar C-stat / d.o.f. values. Based on the data alone it is difficult to make a distinction between the models, although the high $kT$ model fits slightly better. This model seems most likely in the framework of a cavity explosion. Since there is synchrotron emission coming from this part of the remnant, and the forward shock velocity is about 1200 km s$^{-1}$ based on H$\alpha$ emission, the shock must have slowed down due to contact with the cavity wall quite recently, otherwise the electrons would have lost their energy. The cooling timescale of relativistic electrons is $\approx180$ years \citep{helderetal2013}.
If we take that to be the timescale in which dense material got shock-heated, we expect for a density of 1 cm$^{-3}$ 
an ionization age of $5.7\times10^9$ cm$^{-3}$ s. This number agrees quite well with the ionization timescale that we find in this region.

 \begin{table*}
\begin{minipage}{126mm}
\caption{Important diagnostic line fluxes and ratios from different regions of the remnant. The line strengths were obtained using a bremsstrahlung continuum and Gaussians, with an absorption model with N$_{\rmn h}$ fixed at $4\times10^{21}$ cm$^{-2}$. }
\label{tab:lines}

\begin{tabular}{llcccc}

\rule{0pt}{5mm}
& Wavelength (\AA) &  \multicolumn{4}{c}{Flux  $(10^{41}$ ph s$^{-1}$)}  \\
\rule{0pt}{5mm}
& & SW & NW & SE & NE \\
\hline
%\cline{3-6}
%\rule{0pt}{5mm}
\mbox{O\, {\sc vii}}&  $18.60$ & $ 2.17^{+0.10}_{-0.08} $ & $ 0.73^{+0.05}_{-0.05}$ & $1.14^{+0.09}_{-0.08}$ & $0.18^{+0.03}_{-0.00}$ \\
$r$ &$21.60$ & $ 12.60^{+0.56}_{-0.56} $ & $ 5.16^{+0.35}_{-0.34}$ & $6.03^{+0.42}_{-0.43}$ &$ 0.94^{+0.14}_{-0.00}$ \\
$i$ &$21.80$ & $ 2.89^{+0.69}_{-0.57} $ & $ 1.11^{+0.33}_{-0.34}$ &$ 1.23^{+0.51}_{-0.52}$ & $0.00^{+0.21}_{-0.00}$ \\
$f$ &$22.05$ & $ 8.26^{+0.45}_{-0.51} $ & $ 3.35^{+0.34}_{-0.31}$ &$ 2.85^{+0.34}_{-0.34}$ & $0.49^{+0.17}_{-0.00}$ \\
%\rule{0pt}{5mm}
\\
 \mbox{O\, {\sc viii}} &$18.96$ & $ 6.81^{+0.11}_{-0.14} $ & $ 1.73^{+0.09}_{-0.09}$ & $1.66^{+0.14}_{-0.13}$ &$ 0.02^{+0.03}_{-15.01}$ \\
% \rule{0pt}{5mm}
\\
\mbox{Fe {\sc xvii}} &15.01 & $ 1.02^{+0.04}_{-0.03} $ & $ 0.22^{+0.02}_{-0.02}$ & $0.34^{+0.02}_{-0.02}$ & $-$ \\
&15.24 & $ 0.64^{+0.04}_{-0.05} $ & $ 0.16^{+0.02}_{-0.02}$ & $0.21^{+0.03}_{-0.02}$ &$ -$ \\
&16.78 & $ 0.98^{+0.04}_{-0.05} $ & $ 0.21^{+0.03}_{-0.03}$ & $0.24^{+0.04}_{-0.03}$ & $-$ \\
&17.05& $ 1.64^{+0.08}_{-0.06} $ & $ 0.34^{+0.04}_{-0.04}$ &$ 0.60^{+0.06}_{-0.05}$ &$-$ \\
%\rule{0pt}{5mm}
\\
G-ratio$^\dag$\footnotetext{$^\dag$defined as the $\frac{f+i}{r}$ (see text)}
&  & $ 0.88^{+0.08}_{-0.09} $ & $ 0.86^{+0.11}_{-0.11}$ & $0.68^{+0.11}_{-0.11}$ & $0.52^{+0.29}_{-0.19}$ \\
\mbox{O\, {\sc viii}}  / \mbox{O\, {\sc vii}}& & $ 3.14^{+0.15}_{-0.23} $ & $ 2.39^{+0.22}_{-0.21}$ & $1.46^{+0.17}_{-0.15}$ & $0.09^{+0.18}_{-0.09}$ \\
s / C$^\ddag$\footnotetext{$^\ddag$As in \citet{gillaspyetal2011}, defined as s = 16.78 \AA\ + 17.05 \AA, C = 15.01 \AA\ }
& & $ 2.57^{+0.13}_{-0.14} $ & $ 2.50^{+0.32}_{-0.32}$ & $2.48^{+0.26}_{-0.25}$ & $-$ \\
C/ D$^\ast$\footnotetext{$^\ast$C = 15.01 \AA. D = 15.24 \AA} 
& & $ 1.60^{+0.11}_{-0.17} $ & $ 1.41^{+0.25}_{-0.23}$ & $1.64^{+0.23}_{-0.22}$ & $-$  \\

\end{tabular}
%\footnotetext{$^3$ defined as the $\frac{f+i}{r}$ (see text)}
%\footnotetext{$^4$ As in \citet{gillaspyetal2011}, defined as s = 16.78 \AA\ + 17.05 \AA, C = 15.01 \AA\ }
%\footnotetext{$^5$ C = 15.01 \AA. D = 15.24 \AA}
\end{minipage}
\end{table*}

\section{Principal Component Analysis}
\label{sec:pca}

The ejecta composition and distribution is important to identify the nature
of the supernova explosion that created RCW 86 .
One obvious emission feature associated with the ejecta is the Fe-K line at 6.4 keV which  arises in \mbox{Fe {\sc i-xvii}}  \citep*{palmerietal2003}. 
The Fe-K flux per pixel is rather low, so making a map of the Fe-K emission results in a rather noisy image, that is contaminated not
only by background radiation, but also by continuum emission from RCW 86.
However, identifying the components responsible for Fe-K emission is helped by using 
a principal component analysis. This technique reduces the effects of low statistics, because it explores the correlation that exists between the various energy bands,
and is therefore less affected by noise in a single line image.

\begin{figure}
\includegraphics[width=84mm,angle=0]{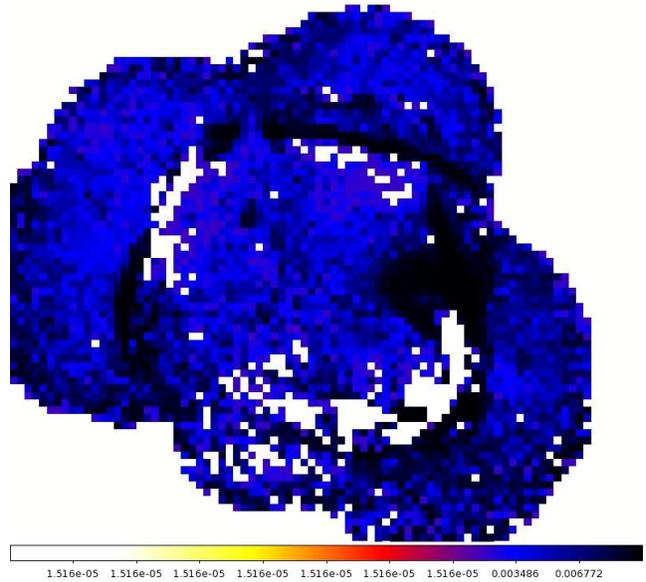}
\caption{Inverted color image of the third principal component (see text).}
\label{fig:PCA_Mode2}
\end{figure}

\begin{figure}
\includegraphics[width=84mm,angle=0]{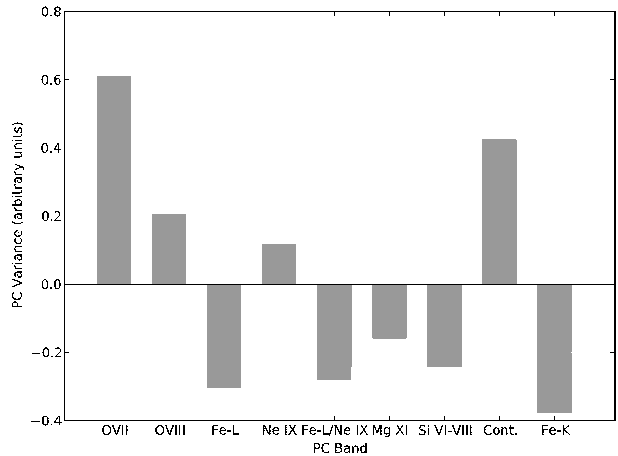}
\caption{PC coefficients of the third principal component. Bands with positive values are correlated in the blue / black regions in Fig. \ref{fig:PCA_Mode2}, while bands with negative values are correlated in the white regions of the same figure. }
\label{fig:eigenvectors}
\end{figure}

\begin{figure}
\resizebox{\hsize}{!}{\includegraphics[angle=0]{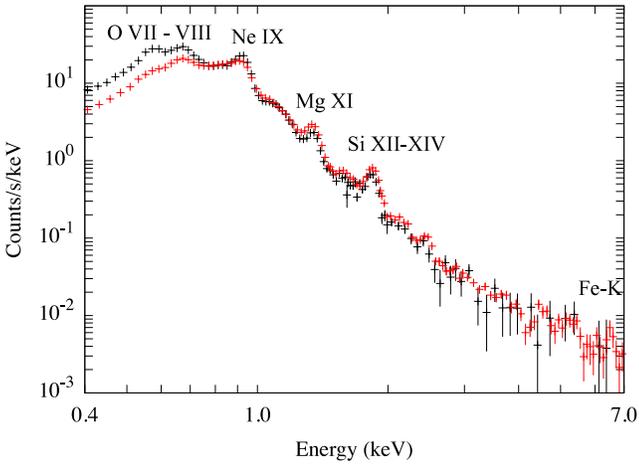}}
\caption{Spectrum of a negative (red) and positive (black) PC score as shown in Fig. \ref{fig:eigenvectors}. In other words, the red spectrum is extracted from a white and the black spectrum from a black region in Fig. \ref{fig:PCA_Mode2}. The correlations implied by the PC coefficients is clearly present in the spectrum, where the red spectrum shows a lot less O and Ne emission, but has a presence of Fe-K, while the black spectrum shows no Fe-K, but has strong O and Ne lines.}
\label{fig:pc2spectrum}
\end{figure}

The PCA gives $n_{\rm bands}$ different components, which all account for part of the variation in the data. As the first PCs account for most of the variation in the original dataset, they often have a clear physical interpretation. For example,
 the second principal component (available in the online material) distinguishes between thermal- and non-thermal emitting regions, which have already been identified using standard analyses. 

The complete results of this analysis are available online. We list here only the most interesting component with regards to Fe-K ejecta distribution, which is the third most significant PC, shown in Fig. \ref{fig:PCA_Mode2}. Fig. \ref{fig:eigenvectors} shows the PC coefficients of this  component. The positive values in Fig. \ref{fig:PCA_Mode2}  shows regions expected to have strong Fe-K and the Fe-L, Mg, and Si bands, but weak hard X-ray continuum, O, and Ne line emission. This is illustrated in the spectrum shown in Fig. \ref{fig:pc2spectrum}, where the red line shows a spectrum taken from the negative regions, while the black line shows a spectrum from the positive regions. The spectrum in black clearly shows more O and Ne, but no Fe-K emission, while the spectrum in red shows clear Fe-K and little O emission. 

The overall morphology of the white region in Fig. \ref{fig:PCA_Mode2} has a striking resemblance to an ellipse with parts in the SE and the West missing. An ellipse shape is more or less the expected emission pattern if the emission is coming from matter distributed as an oblate spheroid shell. If this PC indeed selects for the presence of ejecta, this gives an interesting view on the ejecta distribution. 
As we are primarily interested in Fe-K emission, we checked whether the principal component indeed selects for its presence by making masks of the white regions. We then used these to extract the spectra which are shown in Fig. \ref{fig:FeKspectra}. We fitted the spectra with an absorbed power law to account for the continuum emission, and an absorbed Gaussian with width and centroid as free parameters to model the Fe-K line. The presence of the Fe-K line is significant in the SW, NW and eastern regions. 
The Fe-K line was not significantly detected in the southern part (below the `ellipsoid' shape). 
 
The pointing (0110010701) 
covering the most southern white area in this PC component (see \ref{fig:PCA_Mode2}) has one of the lowest observation times of all pointings,  which could be there reason that Fe-K emission is not significantly detected in this region. We will verify this with planned future {\it XMM-Newton} observations. 

The parameters of the models fitted to the Fe-K spectra are shown in Table \ref{tab:fek}. Several plasma parameters can be determined from the Fe-K line properties. The centroids of the Fe-K line are a diagnostic of the ionization state of the Fe \citep[e.g.][]{palmerietal2003}, where a higher centroid energy corresponds to a higher charge state, which may indicate that the Fe in the eastern part of the remnant has the highest degree of ionization. 
We find that the Fe-K line is broadened in the SW part of the remnant, at a FWHM of $0.20\pm0.03$ keV (or $\sigma = 85\pm13$ eV, corresponding
to $\sigma_{\rm v}=	(4.0\pm 0.6)\times 10^3$~km\,s$^{-1}$). The other spectra show no significant broadening, although this may be due to poorer statistics. The broadening of the Fe-K line in the SW can have different causes: it can arise from a mixture of different ionization states in the plasmas, to different line of sight velocities, or
it may arise from thermal Doppler broadening. We revisit this subject in the discussion. \citet{uenoetal2007} found a broadening of the Fe-K line in the SW part of the remnant of 50 eV. In addition, we confirm their finding of the presence of the K$\beta$ line emission in the SW spectra. Using a gaussian with a width fixed to the K$\alpha$ line width and varying the centroid energy, the line is detected at 4$\sigma$ confidence level, with a line luminosity of $1.93_{-0.51}^{+0.52}\times10^{40}$~ph/s. The centroid is $7.07_{-0.02}^{+0.03}$ keV. Since the probability of a K$\beta$ transition decreases with the number of electrons in the M-shell, the K$\beta$ / K$\alpha$  line ratio is a diagnostic of Fe charge. However, the uncertainty in this ratio of $0.15\pm0.04$ is too large to make a clear distinction between the different charge states.

\begin{figure}
\includegraphics[]{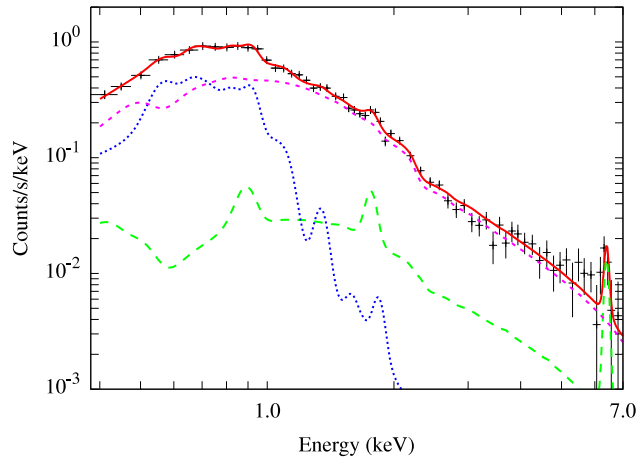}
\caption{EPIC-pn spectrum of the NW region, where there is an ISM plasma component (magenta, small dash), a power law (blue, dotted) and an ejecta component (green, large dash) which shows the Si line at 1.80 keV. This line is mainly produced by \mbox{Si\, {\sc viii-x}}. }
\label{fig:si_spectrum}
\end{figure}

\begin{figure}
\resizebox{\hsize}{!}{\includegraphics[angle=0]{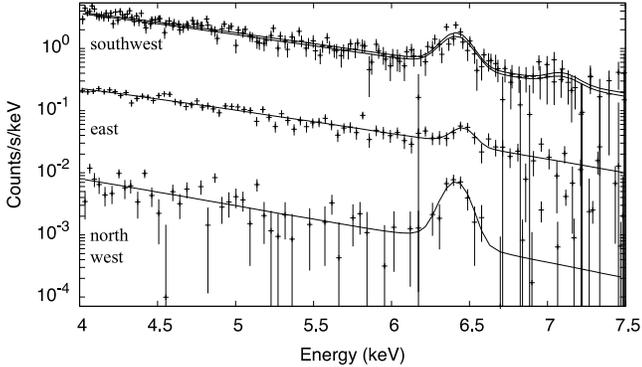}}
\caption{MOS spectra of the Fe-K line of the southwest, east and northwestern region of RCW 86. The MOS 1 and 2 spectra for the east and northwestern regions were added together so the Fe-K line is more visible. }
\label{fig:FeKspectra}
\end{figure}

Besides the presence of ejecta in the form of Fe, there have been no reports so-far of emission from intermediate mass elements (IME) clearly associated with  the 
ejecta in RCW 86, 
such as Si, S, Ar and Ca. We do not detect Ar and Ca emission, but there 
is a significant Si-K line emission present in a region
co-located with the Fe-K emitting plasma, with a low ionization age. Fig. \ref{fig:si_spectrum} shows the spectrum of the white northern inner region of Fig. \ref{fig:PCA_Mode2}. 
We detect the Si-K line emission with both the MOS and pn cameras, but unfortunately the MOS cameras have a strong instrumental Si line at 1.75 keV.
For that reason we concentrated on the pn data for the analysis of the Si-K line emission.  As we have discussed above, using only the Fe-K line emission
we cannot accurately constrain the properties of the Fe plasma for charge states $Z<12$. However, since we find the lowly ionised Si and Fe-K emission at corresponding locations, we are able to fit the Fe and Si with the same NEI model, which allowed us to constrain the ionization age and emission measure of the Fe. This assumes that the Si and Fe emission have been shocked at similar times and locations. In core-collapse supernovae the ejecta are usually fairly mixed 
(i.e. a fair amount of Fe can be found at radii larger than Si, see for example Cas A \citep{hughesetal2000}), but in the case of type Ia supernovae, where ejecta seems stratified \citep*{kosenkoetal2010}, the Si is likely located in a layer outside of Fe and therefore the best fit $\tau$ is an upper limit to the Fe $\tau$. 

The spectrum identified by the third principal component
is best fit by a three component model: one NEI component for the low $kT$ shocked ISM emission in the line of sight, one NEI component to account for low ionised Si and Fe-K, and a power-law component (see Fig. \ref{fig:si_spectrum}). The ISM component cannot account for the Si emission line, and it is therefore 
likely that the Si emission originates form shocked ejecta.
The high $kT$ NEI component likely accounts for the shocked ejecta plasma. 
Notice that the temperature of this component is not well
constrained, but is possibly very high.
Such high temperatures are not unreasonable, as in section \ref{sec:simulations}
we see that the (ion) temperatures of the shocked ejecta can be in excess of 1000 keV. Of course the electron and ion temperatures may be far out of equilibrium.

The ejecta component shows only weak continuum emission, and, therefore, there is a degeneracy in the model between either a higher normalisation coupled with lower Fe abundance, or a lower normalisation coupled with higher Fe abundance. This is caused by the fact that the ejecta consists of almost pure Fe and Si, and contains little hydrogen atoms that can contribute to the bremsstrahlung emission. The continuum is mainly bremsstrahlung caused by electron-Fe collisions. 
To account for the degeneracy we fixed the Fe abundance at 10000, which means that Fe and Si  are the main sources of free electrons. This is well known as a pure Fe plasma (\citet{kosenkoetal2010}, \citet*{vinketal1996}).  
Note that the exact value at which we fix the Fe abundance is not important, as long as it is high enough for the Fe to be the dominant source of electrons. 
The best-fit ionization age of the ejecta component is mainly constrained by the centroid of the Si line, and by the absence of prominent Fe-L emission, which arises at $\tau \geq 2\times10^{9}$ cm$^{-3}$ s. 
Although the uncertainty on some parameters is quite large, the normalisation is reasonably well constrained, and we use this to determine the Fe mass in section \ref{sec:femass}.

\begin{table*}
\begin{minipage}{126mm}
\caption{Fit parameters belonging to the spectrum shown in Fig. \ref{fig:FeKspectra}. }
\label{tab:fek}
\begin{center}
\begin{tabular}{l@{}lcc@{}lc}
 & &  SW & NW & & E \\ 
\hline
power law& $(10^{44}$ ph/s/keV) &$ 0.53_{-0.05}^{+0.07} $  &$ 3.92_{-1.96}^{+6.41} $& $(\times10^{-2})$  & $ 0.21_{-0.02}^{+0.05} $ \\ 
$\Gamma$&                                      &$ 2.92_{-0.07}^{+0.08} $  &$ 3.84_{-0.45}^{+0.64} $&  &$ 3.08_{-0.07}^{+0.13} $ \\ 
Gaussian& $(10^{41}$ ph/s/keV) &$ 1.26_{-0.07}^{+0.07} $  &$ 0.06_{-0.01}^{+0.01}$&  &$ 0.11_{-0.02}^{+0.03} $ \\
centroid& (keV)                                &$ 6.41_{-0.01}^{+0.01} $  &$ 6.41_{-0.01}^{+0.01} $&  &$ 6.44_{-0.02}^{+0.02} $ \\ 
FWHM & (keV)                                 &$ 0.14_{-0.02}^{+0.02} $ &$ 0.001_{-0.001}^{+0.08} $& &$ 0.00_{-0.00}^{+0.13} $ \\ 
\hline
C-stat / d.o.f. &  				  &$ 206 / 159 $ & $357.60 / 139 $& & $ 190.67 / 138 $ \\
\end{tabular}
\end{center}
\end{minipage}
\end{table*}

 \begin{table}
%\begin{minipage}{126mm}
\caption{The parameters of the best-fit model of the northern inner region. }
\label{tab:inner_feK}

\begin{tabular}{lll}
Parameter& Unit & Value \\
\hline
\rule{0pt}{5mm}
\hspace{-0.8mm}ISM & & \\
$N_{\rm h}$ & $10^{21}$ cm$^{-2}$&  $3.60_{-0.14}^{+0.20}$\\
$n_{\rm e}n_{\rm H}V$ & 10$^{55}$ cm$^{-3}$ &$5.17_{-1.52}^{+1.82}$ \\
$kT$ & keV & $0.46_{-0.07}^{+0.05}$\\
$\tau$ & 10$^{9}$ cm$^{-3}$ s & $15.7_{-4.5}^{+6.5}$\\
\rule{0pt}{5mm}
\hspace{-0.8mm}Ejecta & & \\
$n_{\rm e}n_{\rm H}V$ & 10$^{52}$ cm$^{-3}$ &$1.24_{-0.52}^{+1.45}$ \\
$kT$ & keV &$15.0_{-12.1}^{+219}$ \\
$\tau$ & 10$^{9}$ cm$^{-3}$ s &$1.75_{-0.80}^{+1.32}$ \\
Si & &$4156_{-3537}^{+5567}$ \\
Fe &(fixed) &$10000$ \\
\rule{0pt}{5mm}
\hspace{-0.8mm}Power law & & \\
$norm$ &$10^{42}$ ph s$^{-1}$ keV$^{-1}$ &$1.02_{-0.04}^{+0.04}$ \\
$\Gamma$ & & $2.72_{-0.04}^{+0.04}$ \\
\hline
C-stat / d.o.f. & & 230 / 118 \\

\end{tabular}
%\end{minipage}
\end{table}

\section{Simulations}
\label{sec:simulations}

In order to obtain further insight into the structure and
evolution of RCW 86, we perform one dimensional (1D) hydrodynamic simulations.  We retain the idea that the historical remnant was the result of a single degenerate (SD) explosion in a cavity shaped by the mass outflows of the progenitor WD, as argued by  \citet{williamsetal2011}. 
We aim to reproduce its general observational properties,
namely:
a) The remnant has a large diameter of $\sim 29 \pm 6$ pc but at the same time rather low expansion velocities of $\sim 500 - 900~ \rm{km~s^{-1}}$ in the SW/NW and $\sim 1200~ \rm{km~s^{-1}}$ in the SE/NE;
b) the X-ray spectra reveal two emitting plasma components, with the first characterised by a high ionisation age 
and low temperature,  which is associated with the shocked ambient medium (AM), 
whereas the second component is 
characterised by a low ionisation age and high temperatures, which likely corresponds to the shocked ejecta,
and which is responsible for the Fe K line emission around 6.4~keV;
c) 
the Fe-K emission, which traces the hot component, 
is distributed in a shell, relatively close to the forward shock, as indicated by the PCA.

The morphological, dynamical and emission properties of RCW86 deviate substantially from spherical symmetry and 
therefore the remnant as a whole can not be represented well by a single 1D model. Lacking a clear
indication of what the origin is of the asymmetry of RCW 86, we instead model the two extreme parts
in terms of dynamics and emission properties,
the SW and NE regions,  with two separate 1D models.
Nevertheless, we discuss how the asymmetry could have arisen, 
and how the two 1D models together constitutes a reasonable model for the
overall characteristics of the SNR.  
                                   
\subsection{Method}

We employ the AMRVAC hydrodynamic code \citep[][the latter for an example of SNR modeling with AMRVAC]{keppensetal2003,chiotellisetal2013} to simulate the cavity formed by the progenitor system and the evolution of the subsequent supernova explosion in it. We perform our calculation on a one-dimensional grid considering spherical symmetry in the other two. Our simulation size corresponds to $9 \times 10^{19}$~cm (29 pc)
and we use a resolution of 240 shells on the base level. 
Based on the adaptive mesh strategy of the code, we allow for seven refinement levels at which the resolution is doubled wherever large gradients in density or/and energy are present. Hence, the maximum effective resolution becomes $5.9 \times 10^{15}$~cm. 

We simulate the WD's accretion wind \citep[e.g.][]{hachisuetal1996} 
and the formation of the cavity that surrounds the explosion 
centre by creating a mass inflow at the inner boundary of the grid with a density profile of $\rho= \dot{M}/(4 \pi u_w r^2)$ and momentum per unit volume of  $m_r= \rho* u_w$, where $\dot{M}$ is the mass loss rate of the wind, 
$u_w$ is the wind's terminal velocity and $r$ is the radial distance from the source. 
This WD wind cavity simulation  includes four variables: the mass loss rate of the wind, its terminal velocity, the time duration of the wind phase and the density of the interstellar medium (ISM).  
Fortunately, independent studies put constraints on the range of these variables: the mass loss rate and the time of the wind phase should be in agreement with the binary evolution models of the SN Ia progenitors and thus they should be in the range of $10^{-7} -  10^{-6} ~\rm M_{\odot}~yr^{-1}$ and $10^5 - 10^6$~ yr, 
respectively \citep[e.g. Fig 1 in][]{badenesetal2007}, while the wind's terminal velocity is of the order of 
1000 \kms \citep{hachisuetal1996}.
Finally, the density of the ISM  should be around $0.1 - 1 ~\rm cm^{-3}$ to be consistent with the inferred AM densities of RCW86 obtained from infrared observations \citep{williamsetal2011}.

Subsequently, within this wind-blown cavity we introduce the supernova ejecta and we 
let the SNR develop. During the SNR evolution we keep track of the position of its forward shock (FS), contact discontinuity (CD) and reverse shock (RS)  in order to study the dynamical properties of the remnant. 
The initial density and velocity profile of the SN ejecta were taken from the DDTa delayed detonation explosion model 
\citep*{badenes03,bravo96}. 
The explosion model also gives us 
the initial density distribution of Fe and Si ejecta, and we follow their 
distribution during the SNR evolution. In this way we can correlate the results of the simulation with the  Fe  and Si emission properties which the X-ray observations that RCW86 reveals. 
This specific explosion model was chosen, as it produces the maximum mass of iron-group elements and the lowest mass of intermediate mass elements compared to other DDT  and  deflagration models  \citep[see Table 1 of][]{badenes03}. 
Such a chemical composition of the SN ejecta seems to be consistent with what we observe in RCW86 (see also section~\ref{sec:femass}).  
The total Fe and Si mass of the DDTa explosion model is $1.03 ~\rm{M_{\odot}}$ and $0.09~ \rm{M_{\odot}}$ respectively,
and the ejecta mass are that of a Chandrasekhar mass WD ($M_{\rm ej}=1.37$\msun) and the explosion energy
is $1.4 \times 10^{51}$~erg.

\subsection{Results}\label{subsec:hydroresults}

\begin{figure*}%[htbp]       & trim l b r t
\begin{center}$
\begin{array}{ccc}
\includegraphics[trim=0 -50 -50 0,clip=true,width=80mm,angle=0]{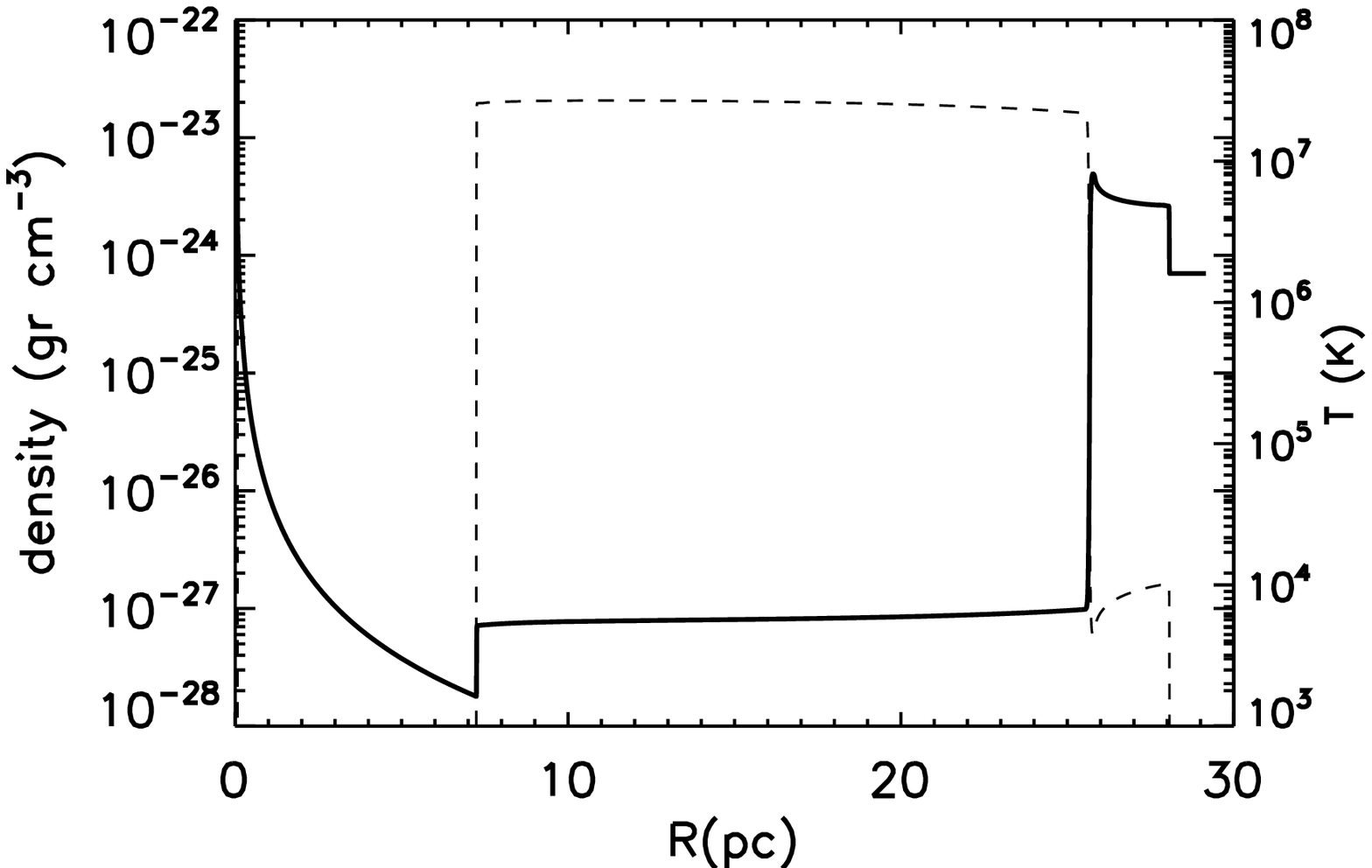} &
\includegraphics[trim=0 -50 -50 0,clip=true,width=80mm,angle=0]{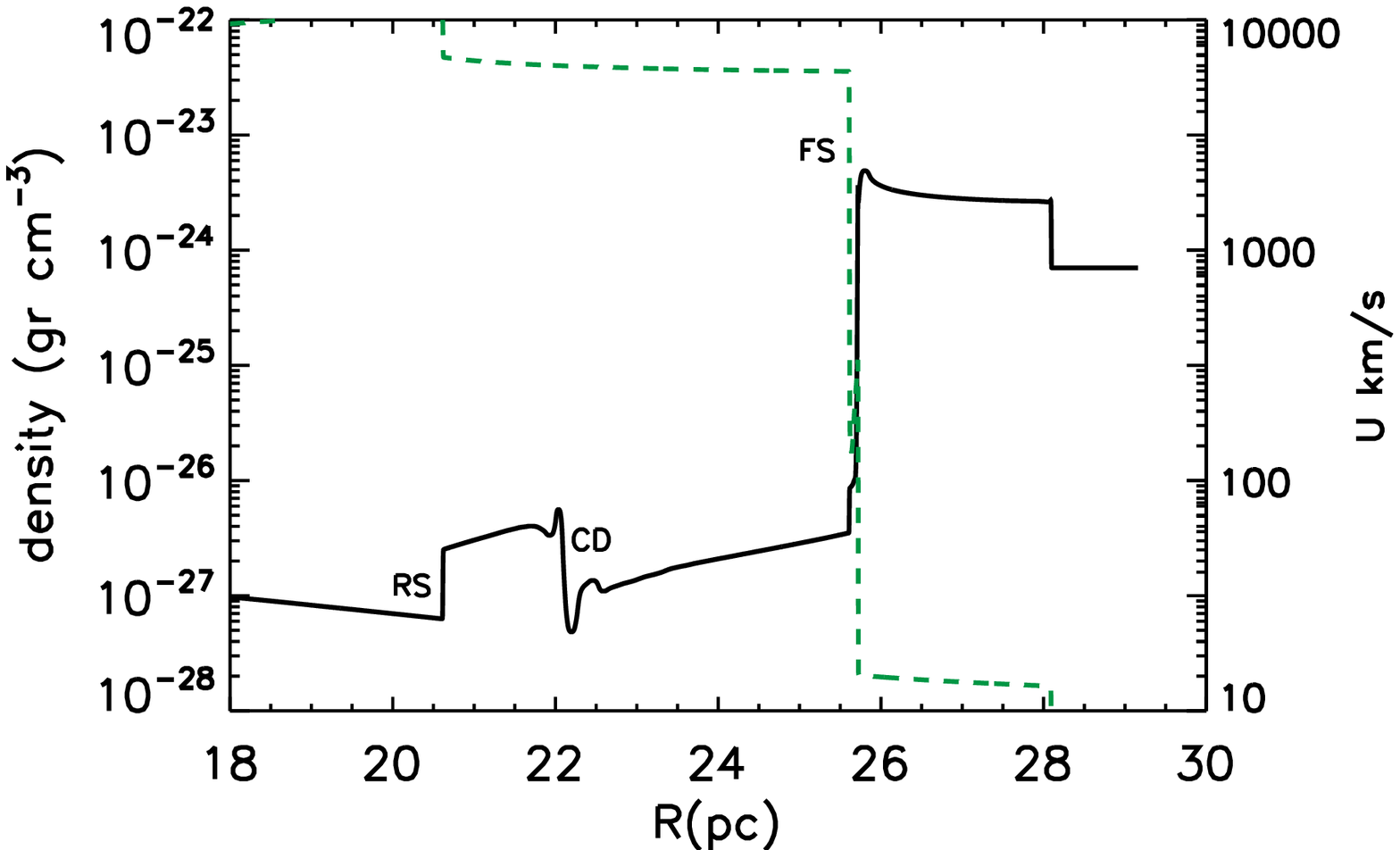}\\
\includegraphics[trim=0 -50 -50 0,clip=true,width=80mm,angle=0]{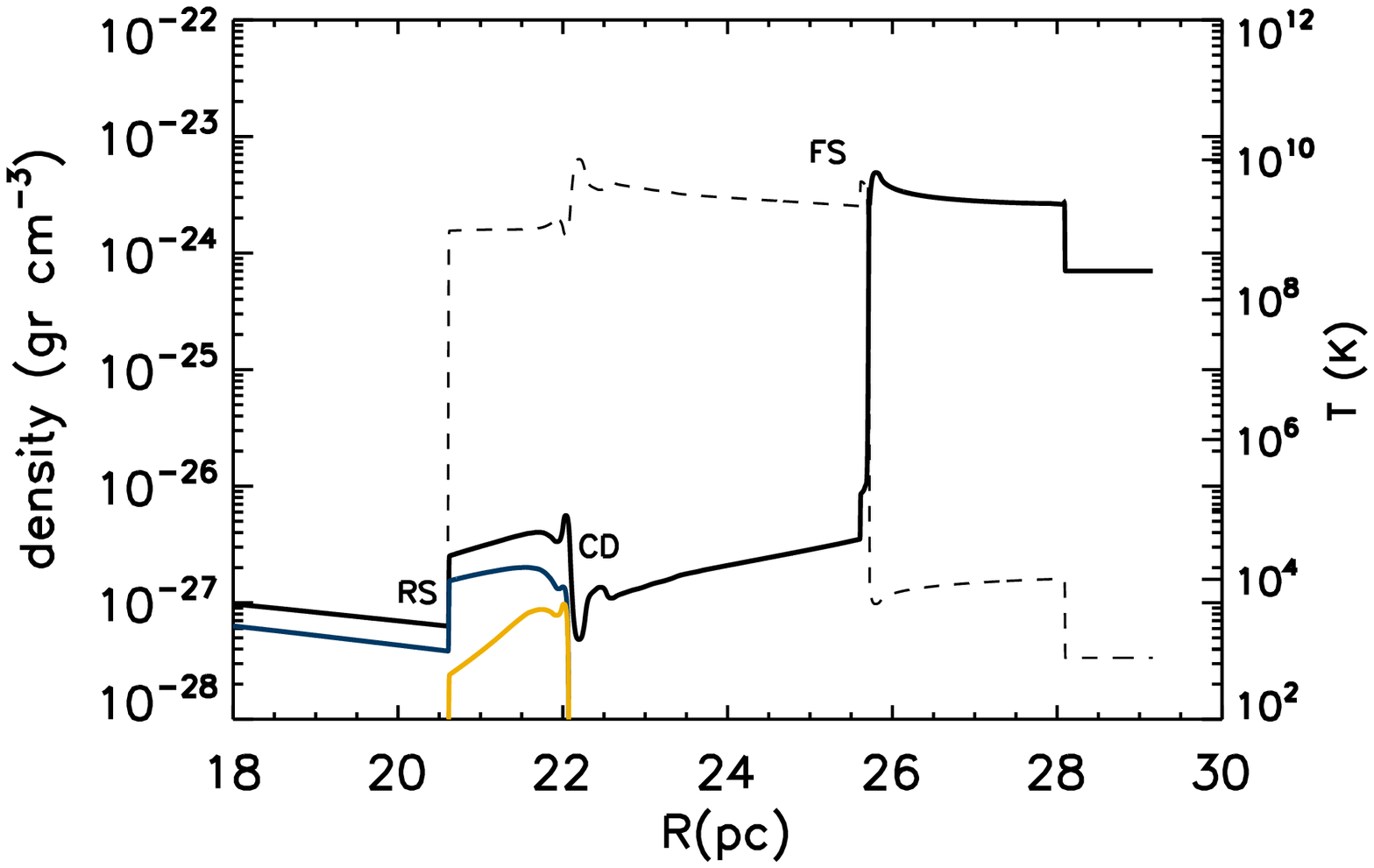} &
\includegraphics[trim=0 -50 -50 0,clip=true,width=69mm,angle=0]{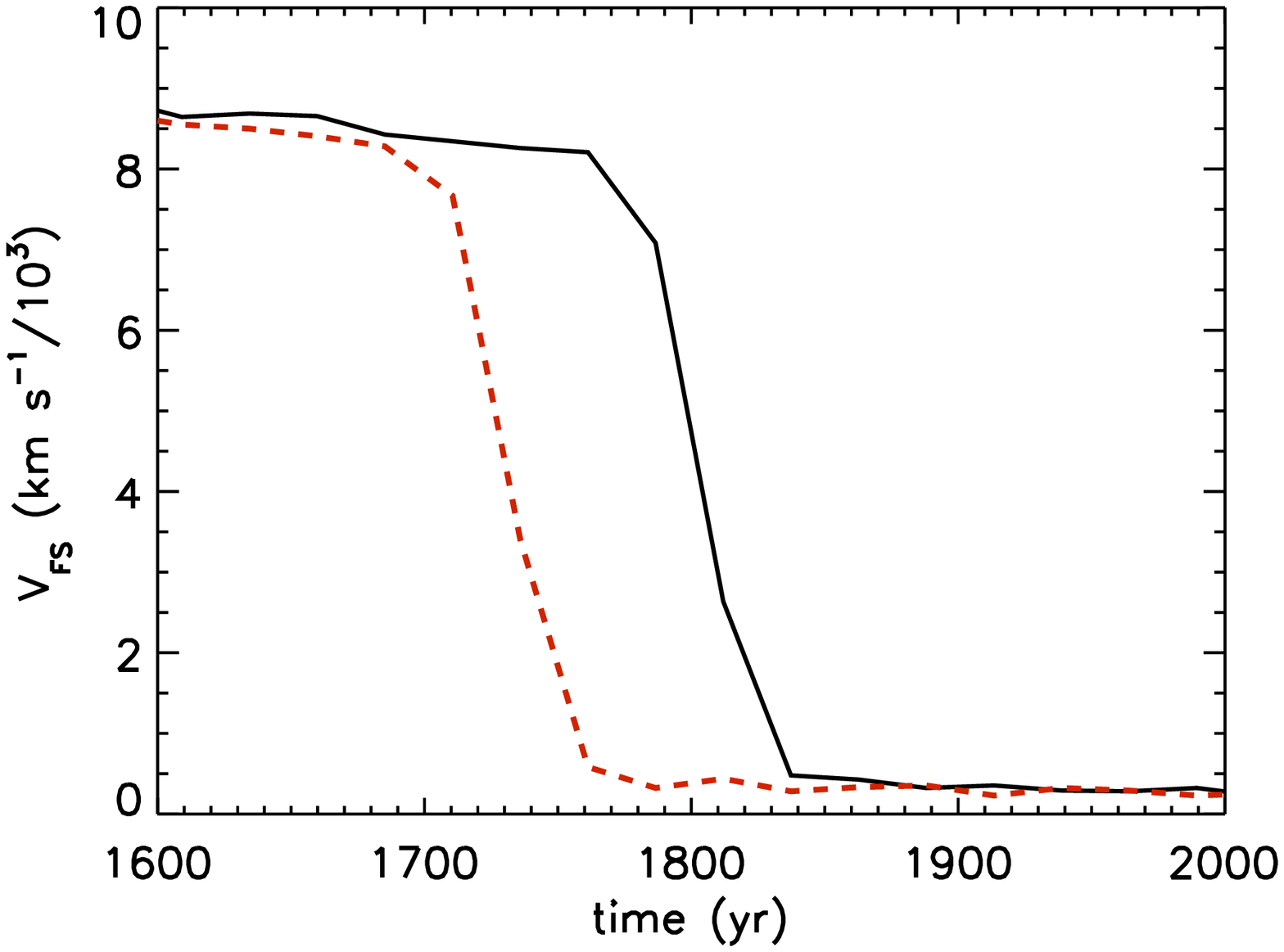} 
\end{array}$
\end{center}
\caption{ Top left: The density (solid line) and temperature (dashed line) radial structure of the wind cavity at 
the moment of the SN explosion, for the situation modeling the NE part of RCW 86.
Top right: The SNR evolving in this cavity at the age of RCW 86, here the
dashed line indicates the {\em plasma} velocity.
Bottom left: The same as the top right panel, but now the dashed
line indicates the plasma temperature, whereas the blue and yellow lines  depict the 
Fe and Si density distribution in the SN ejecta. 
The symbols RS,CD, FS show the position of the reverse shock, contact discontinuity and  forward shock  respectively. 
Bottom right: The time evolution of the FS velocity of the top panel's model (solid/black line) which corresponds to the NE region of the remnant. The red/dashed
line represents a model in which the FS collided with the dense wall 80 years earlier.
This may be a representation of the history of the SE SNR/shell interaction
(see text for details).    
  \label{fig:NEregion}}
\end{figure*}

\subsubsection{NE region}

The NE region of the remnant reveals the highest expansion velocities and at the same time the faintest thermal emission. These properties suggest that the SNR blast wave  interacts with a less dense or/and more recently shocked AM, as compared to the SW region. Intriguingly, this region reveals a discrepancy between the plasma velocity as it is measured from its X-ray and $\rm H \alpha$ emission. In particular, the former shows a high expansion velocity of $6000\pm2800$~\kms \citep{helderetal2009}, which is consistent with the synchrotron emission observed in this region, whereas the $\rm H \alpha$ emission from the NE filaments reveals a much lower mean velocity of $ 1204 \pm 575$~\kms \citep{helderetal2013}. The properties of this SNR portion become more complicated, as the nearby  SE region  reveals a similar  expansion velocity, $1240 \pm 374$ \kms,  but  no  signs of synchrotron emission. 

Given the structure of the ambient medium cavity adopted from our model, the SNR forward shock  can be either inside the cavity, evolving in the low density medium, or it has already reached the edge of the cavity and is currently  propagating in the shocked or unperturbed ISM. Both of these scenarios seem to contradict the observations. On the one hand, if the  forward shock is currently in the cavity, no $\rm H \alpha$ emission is expected as the upstream medium is represented by the hot ($T \sim 10^7$~K) shocked wind (see Fig.\ref{fig:NEregion}, left) and no neutral material is sustained in such high temperatures.  On the other hand, if the forward shock will collide with the density wall of the cavity, the forward shock will substantially decelerate and it will have a low velocity during the rest of the SNR evolution. Therefore, in this case where the SNR has been propagating in the high density ISM for an appreciable time interval the X-ray synchrotron emission observed in the NE region is not expected, as it can only be produced in fast moving shock waves  ($V_s \ge 2000$~\kms).   Furthermore, a prompt collision of the SNR with the density wall  would have triggered a fast moving reflected shock, which would have shocked the ejecta when it was still dense. Thus, in this scenario bright emission and long ionisation ages of the shocked ejecta plasma are expected, something that we  do not observe.

Therefore, it seems that in order to explain at the same time all of the properties of the NE portion of RCW 86, a fine-tuning between these two aforementioned cases is needed. We argue that the scenario that best reproduces the NE observed properties,  is 
one in which the NE part of the SNR is in the transition phase between the two cases described above, and its FS has just recently started to interact with the density wall of the wind cavity. In this case,  the whole evolution history of the SNR's NE region is dominated by its propagation in the cavity. Therefore the resulting NE region is characterised by an extended, low density structure in agreement with the low emissivity and the low ionisation ages observed in this region.   Nevertheless, the recent interaction of the SNR with the edge of the cavity shell resulted in a substantial deceleration of the FS which now is evolving in the rather cold ($T= 10^3 - 10^4$~ K) shocked ISM. 
Therefore, under this scenario, the existence of moderate velocity 
$\rm H \alpha$ filament is also feasible.

Assuming an ISM density of $0.3~\rm cm^{-3}$, 
the cavity size that is consistent with the properties of the NE region
is formed by a wind with a mass loss rate of $1.6 \times 10^{-6}$\msun\ yr$^{-1}$, 
and a terminal wind velocity of 900~\kms, outflowing for 1.0~Myr (Fig. \ref{fig:NEregion}, top left).   
Fig. \ref{fig:NEregion}  illustrates the density and temperature structure of the subsequent SNR at the age of RCW 86 ($t=1830$ yr), as well as the plasma and FS velocity.  The forward shock reaches the edge of the cavity  1760 yr after the explosion having a velocity of $8.5 \times 10^3$ \kms.  After the collision, the FS velocity drops to $\sim 500$~\kms in a time interval less than 60 yr. At the current age of RCW 86 the FS radius is 25.6 pc while its  velocity is $1000$~\kms  in agreement with the $\rm H \alpha$ observations. 
Note that although the shock velocity must have slowed down considerably
in recent times, the plasma velocity inside the SNR retains its high velocity 
for some time, see the top right panel in Fig.~\ref{fig:NEregion}. 
This could explain the high value of the proper motion measured with Chandra 
\citep{helderetal2009}, which is sensitive to a combination of plasma and shock velocity.

Furthermore, 
according to this scenario X-ray synchrotron radiation is possible, produced by relativistic electrons that got accelerated during the recent past of the SNR evolution. Indeed, given the magnetic field of the region of $B \sim 26 \mu$G \citep{aharonianetal2009, vinketal2006}, relativistic electrons of $\sim 100$~ TeV have a synchrotron cooling timescale of 150 - 200 yr. In our model the shock deceleration occurred 70 years ago, therefore  the NE region is expected to still be 
bright in X-ray synchrotron radiation, even though the FS may be slow.  
Note also that given the short time scale for deceleration of the shock
velocity, there may be quite some differences in shock velocities along the NE edge of the remnant, as it is
unlikely that the whole shock encountered the cavity wall at the same time.
Future observations may reveal whether the X-ray synchrotron emission is due to $>$TeV electrons accelerated
in the past, or whether parts of the shock are still fast enough for acceleration to very high energies.

Based on this physical principle, also the transition from the synchrotron emitting NE region to the synchrotron quiescent SE region can be explained by relating the  existence/absence of these non-thermal radiation to the different times where the SNR - density wall collision occurred. Fig. \ref{fig:NEregion} (bottom right) shows two curves of FS velocity evolution: 
The black/solid refers to the aforementioned model in which the SNR/density shell interaction 
took place 1760 yr after the explosion, whereas the red/dashed line represents  
a case where the SNR-wind bubble collision took place at $t = 1680$ yr.
 These two simulations may well represent the 
differences in evolution between the NE part and SE part of RCW 86:
Although the FS velocities at the current age of RCW 86  ($t= 1830$ yr)
have been reported to be rather similar \citep{helderetal2013},
the different times that have passed since the interactions
with the dense shell, may be the reason that the  NE region does display
X-ray synchtrotron emission,  and the NE region shows only thermal X-ray emission.

\begin{figure*}%[htbp]       & trim l b r t
\begin{center}$
\begin{array}{ccc}
\includegraphics[trim=0 0 0 0,clip=false,width=55mm,angle=0]{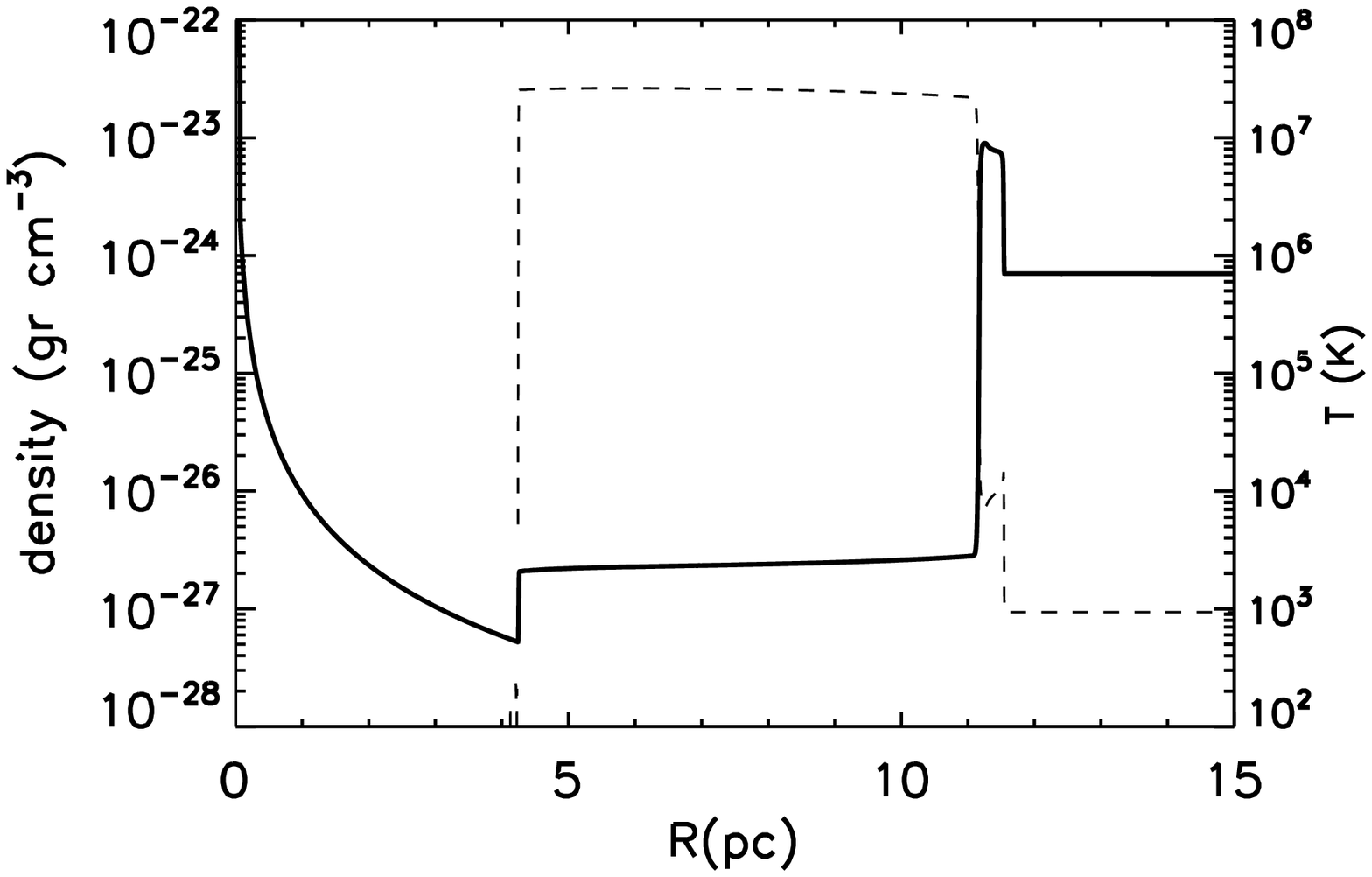} &
\includegraphics[trim=0 0 0 0,clip=false,width=55mm,angle=0]{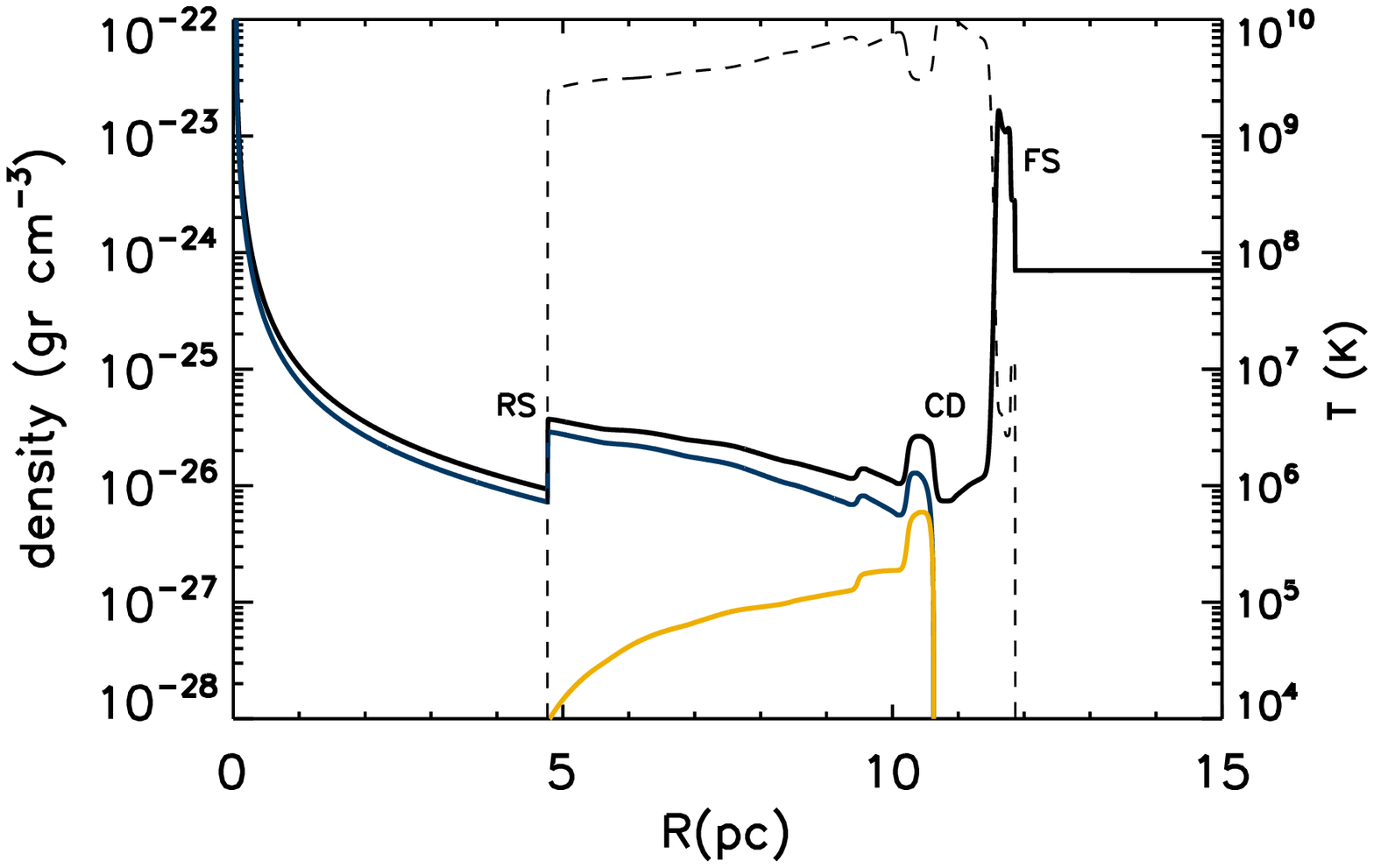} &
\includegraphics[trim=0 0 0 0,clip=false,width=45mm,angle=0]{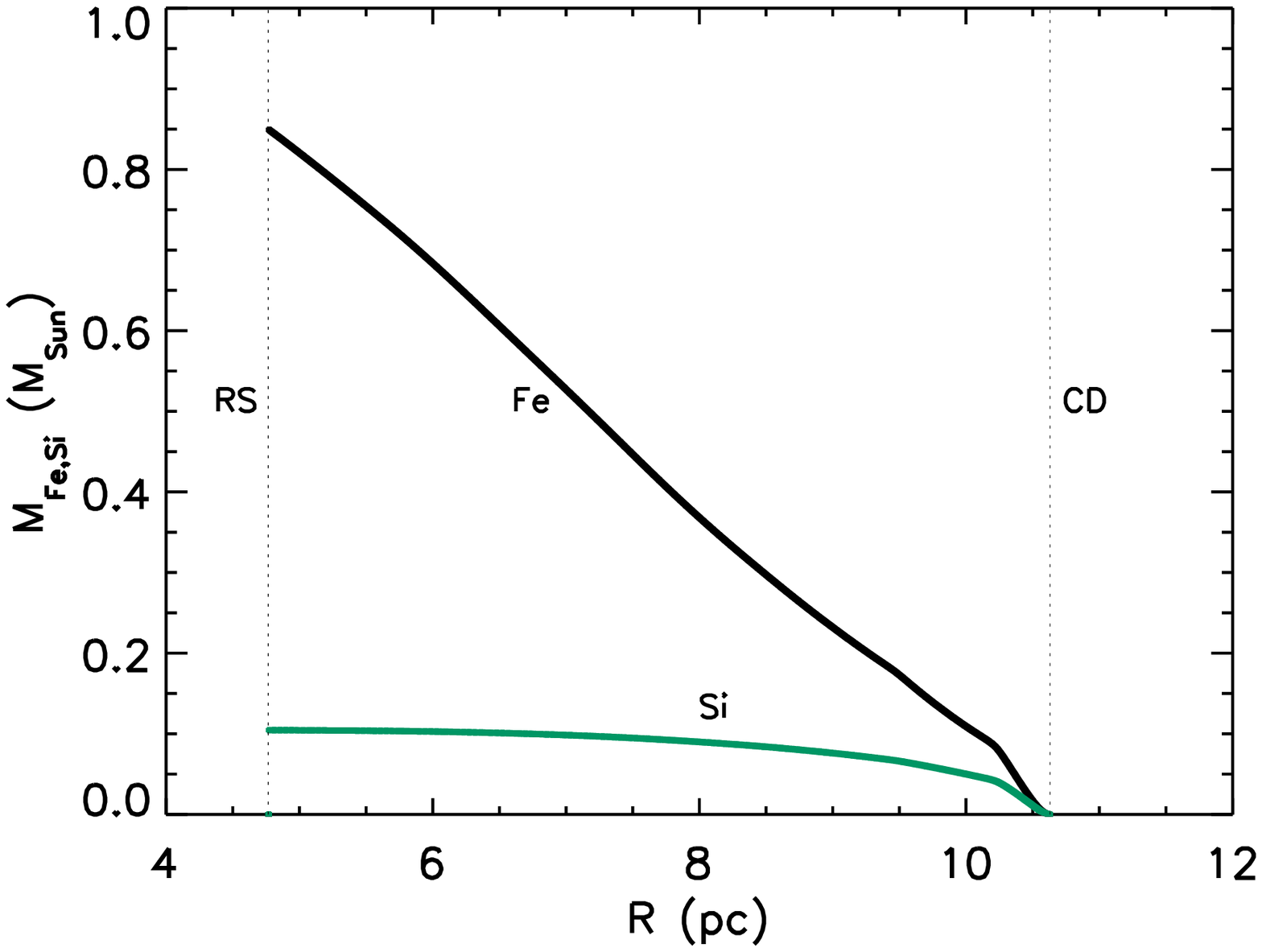} 

\end{array}$
\end{center}
\caption{Left, middle: the same as Fig.~\ref{fig:NEregion} (upper row) but for a model applied for the SW region of RCW 86. Right:  The cumulative mass (from left to right) of the shocked Fe and Si as a function of the SNR radius. The vertical dotted lines correspond to the position of the reverse shock and contact discontinuity. }
  \label{fig:SWdensity}
\end{figure*}

\begin{figure*}%[htbp]       & trim l b r t
\begin{center}$
\begin{array}{ccc}
\includegraphics[trim=0 0 0 0,clip=false,width=55mm,angle=0]{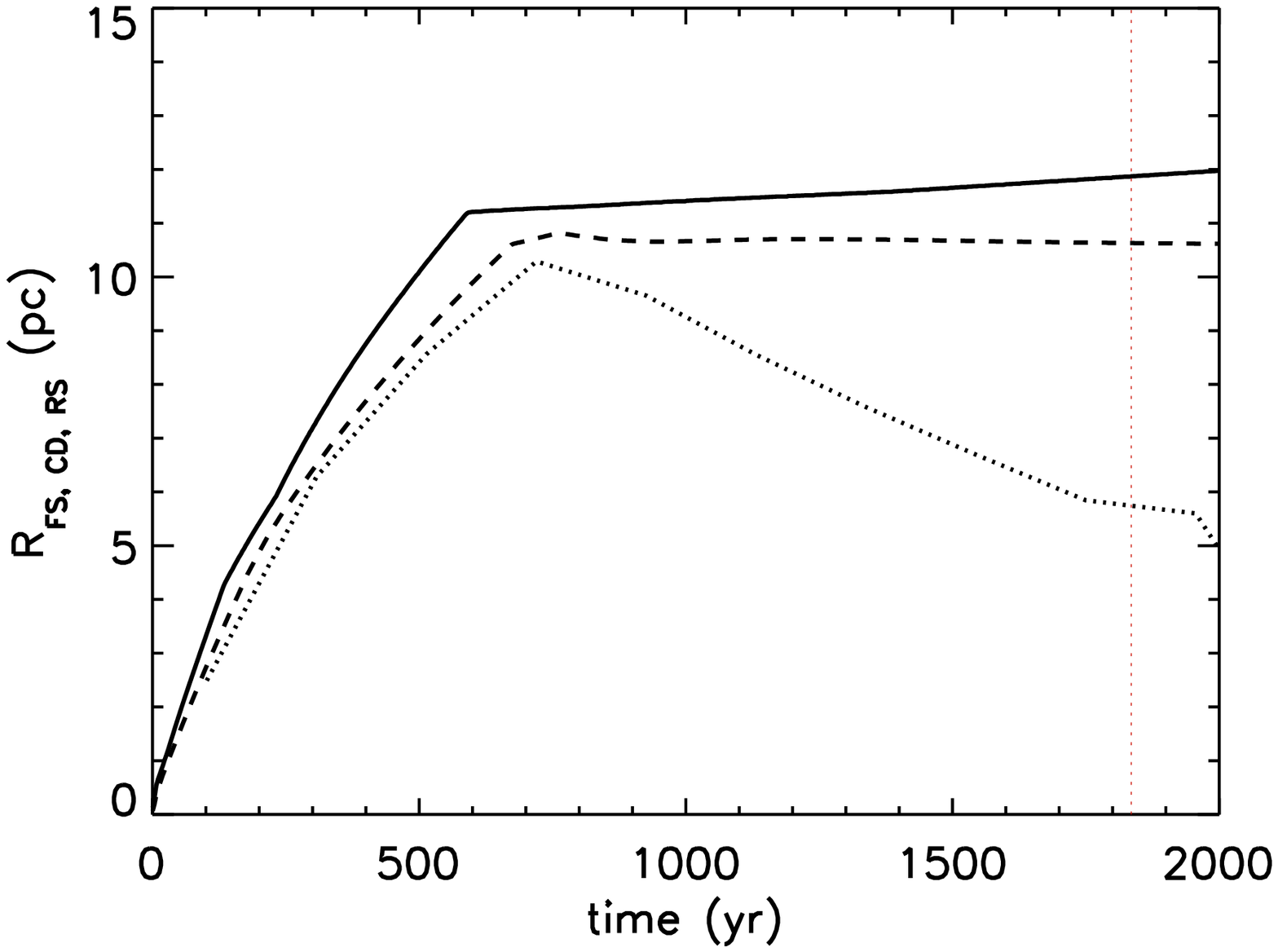} &
\includegraphics[trim=0 0 0 0,clip=false,width=55mm,angle=0]{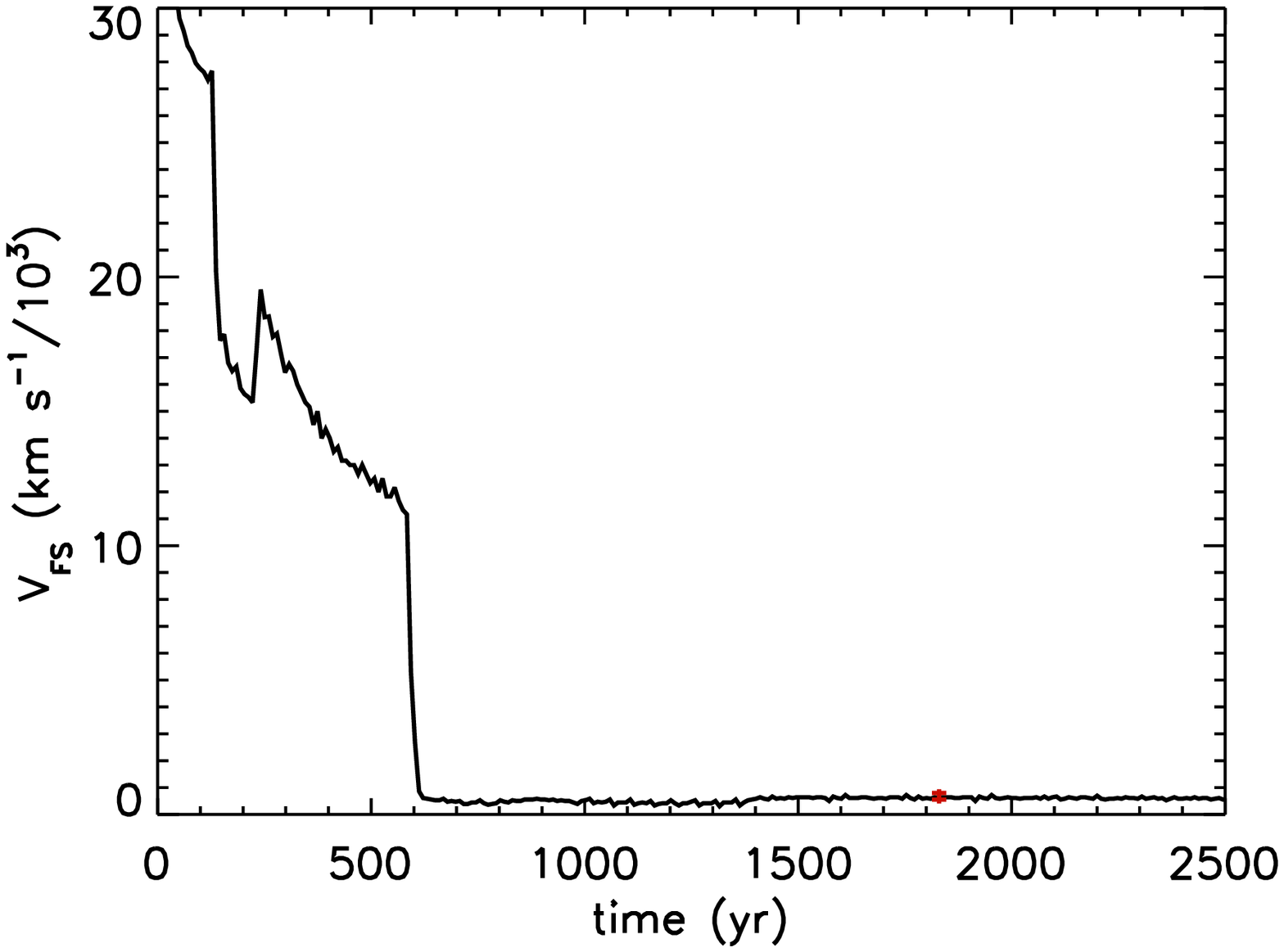} &
\includegraphics[trim=0 0 0 0,clip=false,width=55mm,angle=0]{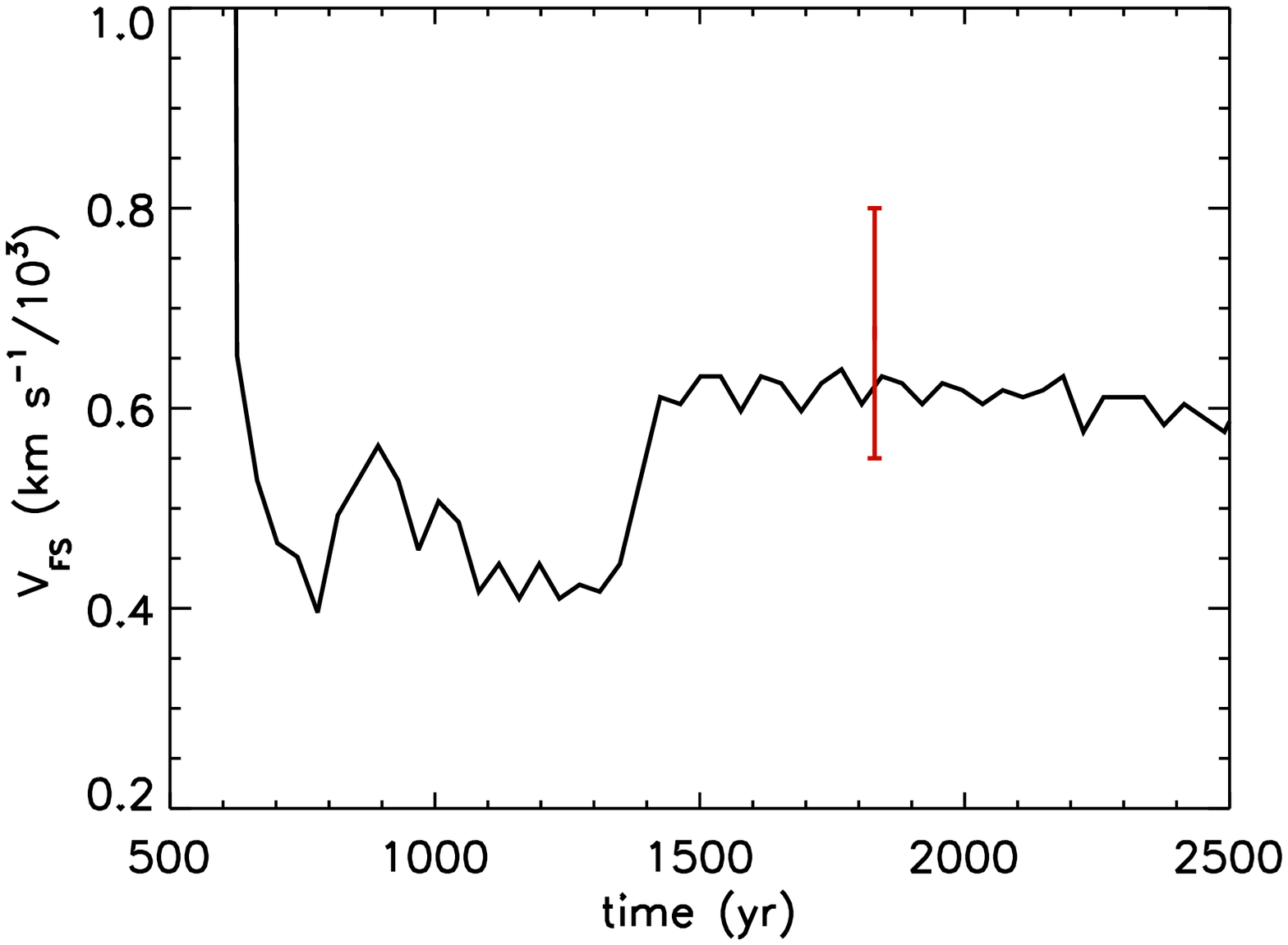} 

\end{array}$
\end{center}
\caption{Left: the time evolution of the forward shock (solid line), the contact discontinuity (dashed line) and reverse shock (dotted line) of the SNR. Middle: the time evolution of the forward shock velocity. Right: same as middle but zoomed in to the latest phase of the SNR. The vertical dashed line marks  the age of RCW 86  }
  \label{fig:SWdynamics}
\end{figure*}

\subsubsection{SW region}

The SW region of RCW 86 is characterised by the highest emissivity and ionisation ages (both of the AM and ejecta 
component), compared to other regions of the remnant.  
This indicates that the shocked AM/ejecta shell in the SW  contains the densest plasma, and/or
was the first to be shocked by the blast wave.
We suggest that the blast wave of the SW portion of the remnant was the first that started to interact with the dense wall of the cavity, resulting in a less extended, but  denser SNR structure. 
This scenario is possible for a non-spherically symmetric wind cavity around the remnant. 
Deviations from spherical symmetry can occur either by a density gradient of the medium around the explosion centre,
 or by a non-zero systemic velocity of the mass losing progenitor system.
Figure  \ref{fig:SWdensity} (left) shows the resulting density structure of the wind cavity for the model which 
best reproduces the properties of the SW region of RCW86. 
In this model we used the same wind parameters as for the NE case, but in order to bring the shocked ISM shell closer to the explosion centre we reduced the wind outflow phase to 0.25 Myr.
  This fast, tenuous, short living wind forms a hot cavity of $\sim 11.5~ \rm pc$ with a density of  $\sim 2 \times 10^{-4} \rm cm^{-3} $. 

Note that invoking a shorter wind phase is an approximation 
needed to model a 2D/3D asymmetry using 1-dimensional
hydrodynamical simulations. For a sketch of how such an asymmetry may arise
see Fig.~7 of \citet{weaver77}. Full modeling of such an asymmetric cavity requires 2D simulations, and taking
into account the proper motion of the progenitor system. In addition, quite some fine tuning is needed to
obtain the right characteristics, as the shape of the cavity depends on the ISM density, wind speed, mass loss rate, wind loss
time scale,  and the proper motion of the system.

The subsequent SNR reaches the density wall 600 years after the explosion and the FS velocity decreases rapidly from $\sim 12000~ \rm km~s^{-1}$  to $\sim 400~ \rm km~s^{-1}$ within 50 yr. 
In the snapshot that corresponds to the age of RCW86 (Fig. \ref{fig:SWdynamics}, right) the FS is propagating  
through
the shocked ISM shell surrounding the wind bubble and has reached a  radius of 12.0 pc. 
Its current velocity is $630~ \rm km~s^{-1}$,  in agreement  with the observations. The CD  at the age of RCW 86  is  at
 a radius of 10.5 pc, which means that Fe and Si rich shocked ejecta can be found up at a radius of  $\sim 90 \%$ of 
the FS radius.  
This result is in agreement with the PCA analysis of RCW 86 which shows that the Fe shell is extended to regions close 
to the outer rim of the SNR (Fig. \ref{fig:PCA_Mode2}).
The RS remain very close to the CD during the whole evolution of SNR inside the cavity. Nevertheless, the collision of 
the SNR with the shocked ISM shell resulted in the formation of a strong, reflected shock, 
which then propagated inward, shocking most of the ejecta (Fig. \ref{fig:SWdynamics}, left). 
At the age of RCW 86, this reflected shock is at a radius of 5.5 pc, having shocked $1.1 \rm M_{\odot}$ of ejecta, 
among which is  $0.85 \rm M_{\odot}$ of Fe  and $0.09 \rm M_{\odot}$ of Si (see Fig. \ref{fig:SWdensity}, right). 
The structure of the SNR at this time is characterised by a very hot ($10^9-10^{10}$~K, Fig.~\ref{fig:SWdynamics}), 
tenuous ejecta surrounded by a much colder but denser shocked AM shell. 
Such a structure is in agreement with the X-ray observations, 
which show that the plasma component that corresponds to the shocked AM reveals a higher ionisation age and 
lower temperature compared to the shocked ejecta plasma. 
The shocked ejecta plasma is very hot, which may be the reason that the Fe-K emission is intrinsically broadened
as a result of thermal Doppler broadening. The reported width for the Fe-K emission in the SW of RCW 86 is
$\sigma_{\rm v}=(4\pm 0.6)\times 10^3$~km/s, which suggests Fe temperature of $kT=m_{\rm Fe}\sigma_{\rm v}^2=10\pm 3$~MeV 
or $1.1\pm 0.3\times 10^{11}$~K. This corresponds to an equivalent hydrogen temperature of $2\times 10^9$~K, if
one assumes non-equilibration of temperatures. 
It is this equivalent hydrogen temperature that is indicated in Fig.~\ref{fig:NEregion} and \ref{fig:SWdensity}.
This model could be further tested with high resolution X-ray spectroscopy made possible with the Astro-H satellite
\citep{takahashietal2013}.

\section{Discussion}

\subsection{Fe mass in RCW 86}
\label{sec:femass}

A clear difference between remnants of core collapse and type Ia supernovae is the amount of Fe present in the remnant \citep{vink2012}.
Core collapse supernovae typically produce less than 0.1~\msun\ of iron, whereas type Ia supernovae produce 0.5-1 \msun\ of iron.
The amount of Fe present in RCW 86 is therefore important to determine the nature of the supernova explosion.
With the parameters obtained in section \ref{sec:pca} there are two ways to determine the mass of Fe in the oblate spheroid shell. The first method uses the emission measure $Y = n_{\rm e}n_{\rm H}V$. This can be converted to an iron mass using two parameters that can be extracted from SPEX,  $n_{\rm e} / n_{\rm H}$ and $n_{\rm Fe} / n_{\rm H}$, as follows \citep[e.g.][]{kosenkoetal2010}:
\begin{equation}
M_{\rm Fe} = \sqrt{\left(\frac{Y}{n_{\rm e}/n_{\rm H}} \left( \frac{n_{\rm Fe}}{n_{\rm H}}\right)^2 \frac{1}{V_{\rm x}}\right)} \times 56m_{\rm p}\times V_{\rm shell} 
\end{equation}
The part underneath the square root equals the number density of Fe particles $n_{\rm Fe}$. $V_{\rm x}$ is the volume corresponding to the region from which the spectrum is extracted, $m_{\rm p}$ is the proton mass and $V_{\rm shell}$ is the volume of the oblate spheroid shell in which we assume the shocked Fe ejecta is distributed. Using a distance of 2.5 kpc, $V_{\rm x} = 1.2\times10^{57}$ cm$^{3}$ and $V_{\rm shell} = 7.1\times10^{58}$ cm$^{3}$. Using the parameters listed in table \ref{tab:inner_feK} we obtain from SPEX: $n_{\rm e} / n_{\rm H} = 5.78$ and $n_{\rm Fe} / n_{\rm H} = 0.326$. This leads to $M_{\rm Fe} = 1.4^{+0.7}_{-0.3}M_{\odot}$. This value is very close to the amount of Fe that is expected in type Ia explosion models. The assumptions we used here are that the model we use to describe the ejecta is correct, and that the ejecta are distributed as an oblate spheroid shell described earlier. 

The second method involves using the ionization age $\tau = n_{\rm e}t$. $M_{\rm Fe}$ can be written as:
\begin{equation}
 n_{\rm Fe} = \frac{M_{\rm Fe}}{V_{\rm shell}\,56\,m_{\rm p}},
\end{equation}
where $V_{\rm s}$ is again the volume corresponding to the oblate spheroid shell. From $n_{\rm e} / n_{\rm H}$ and $n_{\rm Fe} / n_{\rm H}$ listed above we obtain $n_{\rm Fe} =  n_{\rm e} / 17.7$, so that: 
\begin{equation}
M_{\rm Fe} = \frac{n_{\rm e}t}{t} \frac{56m_{\rm p}\times V_{\rm shell}}{17.7}.
\end{equation}
Using $n_{\rm e}t = 1.75\times10^{9}$ cm$^{-3}$, $t = 1000$ yr and $V_{\rm shell}$ listed above, we obtain $M_{\rm Fe} = 10.3M_{\odot} $.
This calculation assumes a plasma where Fe ions are the sole source of free electrons. This is then the Fe mass needed to reach an ionization age of $1.75\times10^9$ cm$^{-3}$ s in a time t. The amount of Fe is too large for either a type Ia or CC progenitor for RCW 86. However, this method has a number of uncertainties. 
First of all, $\tau\equiv \int n_{\rm e}(t)dt$ is a time integrated quantity.
Using it to estimate the current value of 
$n_{\rm e}$ introduces uncertainty, because at the time it was shocked the plasma had a higher density than the present day density. On the other hand the electron density $n_e$ that we use is based on the current ionization state of the plasma, which was less ionized in the past.
Secondly, Fe is not the sole source of electrons in the shell of ejecta, as Si also contributes greatly. This leads to an overestimation of the mass with a factor of $\sim1.6$. Finally, there is an uncertainty in the shock age.  

In the calculations above we assumed that the particles are distributed homogeneously in space, while matter in SNR ejecta in reality is probably clumped \citep[e.g.][]{orlandoetal2012}. If the matter is clumped, this means that the calculations above are an overestimation of the true Fe mass, for the emissivity and $n_{\rm e}t$ are higher when clumped, for the same amount of particles.
This clumping can be approximated by assuming a filling fraction $f$, so that the matter is effectively distributed over a smaller volume $fV$. The emission measure and the $n_{\rm e}t$ method have a square root and a linear dependance on $f$, respectively. Using a reasonable filling fraction $f=0.1$ brings the mass estimates much closer, at  $0.44(f_{0.1})^{0.5}$ \msun\ and $1.03f_{0.1}$ \msun. In addition, using the emission measure method above with the same filling fraction, we can calculate $M_{\rm Si, shell}=0.15(f_{0.1})^{0.5}$ \msun.
Assuming constant density from the Fe ejecta shell to the centre, $M_{\rm Fe, total} \simeq 1~(f_{0.1})^{0.5}$ \msun.
The high Fe mass that we find supports the idea that RCW 86 is the result of a type Ia explosion.

\subsection{Progenitor type}

We have shown that the Fe mass of RCW 86 is indeed  
consistent with a thermonuclear explosion, strengthening the case that
RCW 86 is the result of a type Ia explosion.
In addition, we have shown that the wind-blown cavity 
in which the SNR evolves is well reproduced
by a continuous outflow with a mass loss rate of $1.6 \times 10^{-6}$ \msunyr, 
a wind terminal velocity of 900 \kms, which has been outflowing from the progenitor system for 1~Myr. 
In the type Ia regime, these fast moving, tenuous winds can only emanate 
from the WD's surface for such a long time in case 
of a single degenerate progenitor system. 
These mass outflows are known as accretion winds and are thought to 
accompany mass transfer processes that are characterised by 
high accretion rates ($\dot{M} \ge 10^{-7} - 10^{-6}$  \msunyr, depending on the WD mass; \citet{hachisuetal1996}; \citet*{hachisuetal1999}).

As shown by \citet{badenesetal2007}, if indeed RCW 86 is expanding in a wind blown cavity, then it is the only known type Ia SNR for which the dynamical and emission properties are in agreement with the progenitor models that predict these accretion wind outflows. 
The existence of accretion winds is crucial 
for a better understanding of how single degenerates systems contribute
to the observed type Ia supernova rate.
The reason is that the accretion wind is the only suggested process
that can regulate the actual accretion onto the white dwarf, thereby ensuring
a steady, stable mass transfer.

Based on our hydrodynamical models, the wind cavity contains  1.6 \msun\ of material ejected from the progenitor system. This, in combination with the fact that at least 0.3 solar masses of material should have been accreted to the WD in order to reach the Chandrasekhar mass (the maximum mass of a CO WD at its creation is 1.1 \msun), makes a total donor star envelope mass deposit of 1.9 \msun. Based on the semi empirical WD initial-final mass relationship  \citep{salarisetal2009} our model predicts that the minimum mass of the donor star is 2.6 \msun. 
Population synthesis models show that donor stars with these masses are within 
the binary parameter space  that leads to type Ia explosions \citep[e.g.][]{hanpodsiadlowski2004}. We also showed that the progenitor system had a $\sim30\%$ offset with respect to the geometrical centre in the SW direction. However, given the large number of variables and simplification that we used in our modelling, these predictions have a high level of uncertainty, and further 2D modelling is needed.

Alternative mass outflows from single degenerate SNe Ia are also possible to form a cavity around the explosion center. These outflows could be the wind of a red giant donor star or successive nova explosions \citep{chiotellisphdthesis}.   However, all of these processes most likely fail to provide the energy budget needed to excavate a cavity with diameter $\sim 30 $ pc. 
In our modelling, the cavity which was able to reproduce the properties of
RCW 86 was formed by a wind outflow with mechanical luminosity of $5 \times 10^{35} \rm erg~s^{-1}$  emanating for 1 Myr, depositing a total energy
of $1.5 \times 10^{49}$~erg. 
Red-giant winds have typical mass loss rates of $10^{-6} - 10^{-7}$ \msunyr\  
and have wind terminal velocities of $\sim 50 - 70$ \kms.
The lifetimes of the red-giant phase are 0.1 - 1 Myr. 
As a result, red-giant wind luminosities are $\sim 10^2 - 10^4$ smaller than 
provided by the accretion wind used for our model. 
Based on analytical models of wind dynamics we find that red-giant winds 
can form a cavity with maximum radii of $\sim 1 - 10$~pc  
for an ISM density of $0.3~\rm cm^{-3}$ \citep[see eq. 4.2 of][]{koomckee92}. 
Such cavities are rather small compared to the large radius of RCW 86, 
but cannot be completely excluded given the uncertainties in the
AM densities.
As for cavities created by nova explosions,
a typical energy of a nova explosion is $2 \times 10^{44}$~erg 
\citep{orlando12}. In order to 
provide enough energy to blow a cavity the size of RCW 86,
$10^5$  nova explosions are needed prior to the type Ia explosion.
This number of novae is too large as compared to binary evolution model 
predictions \citep[see e.g.][]{hachisu08}.

Interestingly, the formation of low densities cavities are also predicted from the double degenerate regime. These can be either a wind cavity shaped by continuum radiation driven winds  during the thermal evolution of the merger \citep{shen12} or a planetary nebula formed by the envelope ejection of the secondary star \citep*{kashisoker2011}. Nevertheless, even if  both scenarios predict  similar wind properties as those used in our model ($\dot{M} \sim 10^{-7} - 10^{-6}$ \msunyr, $u_w \sim 10^3$ \kms) they fail to reproduce the CSM around RCW 86 due to the timescales evolved. 
In both cases the fast wind lifetime is of the order of $10^4$ yr. 
During such a small time, the mass outflow cannot excavate the extended cavity that surrounds RCW 86.  
 Using again the predictions of \citet{koomckee92} and an AM  density of  $0.3~\rm cm^{-3}$ we find that the typical radius of 
the cavity is up to 2~pc for WD merger winds, 
while for planetary nebulae the cavities are expected to be lower, 
as the fast wind first has to sweep up the dense material
from a previous evolutionary phase.  
Observational evidence supports this conclusion, as 
the typical sizes of planetary nebulae are $\sim 0.2$ pc %\citep{kwok94}. 

\section{Conclusion}

We presented the most complete X-ray view of RCW 86 so far, using all {\it XMM-Newton} pointings currently available. We fitted the combined RGS and MOS spectra of four quadrants of the remnant, thus obtaining detailed plasma parameters of both the shocked ambient medium and ejecta plasma components for a large part of the remnant. The large differences in ionization ages between the shocked ejecta and shocked ISM are most naturally explained by a supernova exploding in a wind-blown cavity, where the reverse shock has been close to the forward shock for a large part of the lifetime of the remnant so that the ejecta have substantially lower ionization ages compared to the shocked ISM. From the ambient medium ionization ages, we can construct an interaction history of the forward shock with the cavity wall, for which we find that the SW has been shocked earliest, followed by the NW, SE and finally the NE. The NE part of the remnant may have just started to interact with the cavity wall, which could explain the presence of synchrotron emission at he FW shock in this region while the H$\alpha$ shock velocity is $\approx1200$ km s$^{-1}$. Using principal component analysis, we obtained the highest resolution map of the location of ejecta emission (most prominently Fe-K), thus far. The ejecta seem located in an oblate spherical shell,  close to the forward shock. We obtain an Fe mass of around $1~ (f_{0.1})^{0.5}$ M$_{\odot}$, consistent with a type Ia progenitor.

In addition, we used hydrodynamical simulations to show that the current size and  dynamical and spectral parameters of RCW 86 can be well-reproduced by a white dwarf exploding in a wind-blown cavity, as suggested by \citet{badenesetal2007, williamsetal2011}. Our work further strengthens the notion that RCW 86 had a single degenerate progenitor system, which actively modified its environment.

\section{Acknowledgements}
S.B. is supported financially by NWO, the Netherlands Organisation for Scientific Research. The results presented are based on observations obtained with {\it XMM-Newton}, an ESA science mission with instruments and contributions directly funded by ESA Member States and the USA (NASA).  We would like to thank prof. dr. R.A.M.J. Wijers and dr. O. Pols for carefully reading the manuscript, prof. dr. J. Raymond for pointing us to the Vedder et al. (1986) paper, C. Abate for interesting discussions on stellar evolution and dr. C. Badenes and dr. E. Bravo for providing the DDTa explosion model used for the hydrodynamical simulations.

\bibliographystyle{mn2e}
\bibliography{sjors}

\end{document}